\documentclass[twocolumn]{aastex63}
\shorttitle{The PHOENIX Exoplanet Retrieval Algorithm: PETRA}
\shortauthors{Lothringer et al.}

\usepackage{amsmath}
\usepackage{xspace}
\usepackage{float}

\newcommand{\microns}{$\mu$m}
\newcommand{\chisq}{$\chi^2$}

\newcommand{\spitzer}{{\it Spitzer}}

\begin{document}
\title{The PHOENIX Exoplanet Retrieval Algorithm and Using H$^{-}$ Opacity as a Probe in Ultra-hot Jupiters}
\author[0000-0003-3667-8633]{Joshua D. Lothringer}
\affiliation{Department of Physics and Astronomy, Johns Hopkins University, Baltimore, MD, USA}
\affiliation{Lunar and Planetary Laboratory, University of Arizona, Tucson, AZ, USA}

\author[0000-0002-7129-3002]{Travis S. Barman}
\affiliation{Lunar and Planetary Laboratory, University of Arizona, Tucson, AZ, USA}
\vspace{0.5\baselineskip}
\date{\today}
\email{jlothri1@jhu.edu}

\begin{abstract}

Atmospheric retrievals are now a standard tool to analyze observations of exoplanet atmospheres. This data-driven approach quantitatively compares atmospheric models to observations in order to estimate atmospheric properties and their uncertainties. In this paper, we introduce a new retrieval package, the PHOENIX Exoplanet Retrieval Analysis (PETRA). PETRA places the PHOENIX atmosphere model in a retrieval framework, allowing us to combine the strengths of a well-tested and widely-used atmosphere model with the advantages of retrieval algorithms. We validate PETRA by retrieving on simulated data for which the true atmospheric state is known. We also show that PETRA can successfully reproduce results from previously published retrievals of WASP-43b and HD 209458b. For the WASP-43b results, we show the effect that different line lists and line profile treatments have on the retrieved atmospheric properties. Lastly, we describe a novel technique for retrieving the temperature structure and $e^{-}$ density in ultra-hot Jupiters using H$^{-}$ opacity, allowing us to probe atmospheres devoid of \added{most }molecular features with JWST.

\end{abstract}

\keywords{planets and satellites: atmospheres, methods: numerical}

\section{Introduction}

Retrieval algorithms are now widely-used to infer atmospheric properties, like the composition and temperature structure, from observations of sub-stellar objects. Retrieval algorithms have two basic parts: a forward model that produces a spectrum and a statistical framework that chooses parameters for that forward model and compares the spectra with observations. A primary advantage of using retrieval algorithms over grid-based searches is that retrievals provide robust estimations of parameter uncertainties, correlations, and degeneracies through efficient sampling of parameter space and the posterior distribution. While retrieval forward models are generally not fully self-consistent and require multiple parameterizations and assumptions for the sake of computational efficiency, retrievals are a valuable method to interpret observations.

Retrievals have been used to analyze transit \citep[e.g.,][]{line:2012,benneke:2012,waldmann:2015,barstow:2017,macdonald:2017,howe:2017,molliere:2019} and secondary-eclipse observations \citep[e.g.,][]{line:2013b,gandhi:2018,waldmann:2015,evans:2017,molliere:2019,kitzmann:2020,himes:2020}, as well as  observations of self-luminous objects like directly-imaged exoplanets \citep{lee:2013,lavie:2017,gravity:2020} and brown dwarfs \citep{line:2015,line:2017,burningham:2017}. Recently, the application of retrieval algorithms to combine low- and high-resolution data has been explored \citep{brogi:2017,brogi:2018,fisher:2019,gandhi:2019b,gibson:2020}. See \cite{madhu:2018review} for an overview of \replaced{most}{many} existing retrieval algorithms\added{ and \cite{barstow:2020b} for a discussion of open problems in retrieval analysis}.

PHOENIX is a well-tested self-consistent atmosphere model that has been used to study the atmospheres of stellar and sub-stellar atmospheres for decades \citep{hauschildt:1997,hauschildt:1999,allard:2010,barman:2001,barman:2011,lothringer:2018b,lothringer:2019}. In self-consistent frameworks, the model is typically iterated until certain convergence criteria are met (i.e., radiative-convective equilibrium). Self-consistent models provide us with our best prediction of the structure and composition of an atmosphere based on the physical assumptions included in the model. \added{The comparison of observations with self-consistent models can provide insight into the processes at work in the atmosphere.}

In this work, we introduce a new retrieval framework, the PHOENIX ExoplaneT Retrieval Algorithm (PETRA), which utilizes PHOENIX as its forward model. PETRA allows us to combine many of the strengths of one of the most widely-used atmosphere models with the advantages of retrieval algorithms. PETRA's use of PHOENIX's opacity database will prove useful in the identification and characterization of molecules and atoms in complex exoplanet atmospheres, as well as in understanding line list biases. PHOENIX's line sampling methods will also be effective at retrieving atmospheric properties from high-spectral resolution observations. Additionally, PHOENIX's broad applicability is advantageous in a retrieval forward model, as we will be able to use PETRA to explore exoplanet, brown dwarf, and stellar atmospheres in different geometries while being able to compare to self-consistent predictions from the same model.

Section~\ref{sec:methods} explains the structure of PETRA, as well as the parameterizations and statistical framework used. In Section~\ref{sec:tests}, we validate PETRA by presenting retrievals on simulated data with known atmospheric parameters. We also compare PETRA to other retrieval tools by comparing results for WASP-43b and HD 209458b. Lastly, we demonstrate a novel use of PETRA to retrieve the temperature structure and $e^-$ density in ultra-hot Jupiters from James Webb Space Telescope (JWST) simulated observations using H$^-$ opacity in Section~\ref{sec:hminus}.

\begin{figure}
	\center    \includegraphics[width=3.4in]{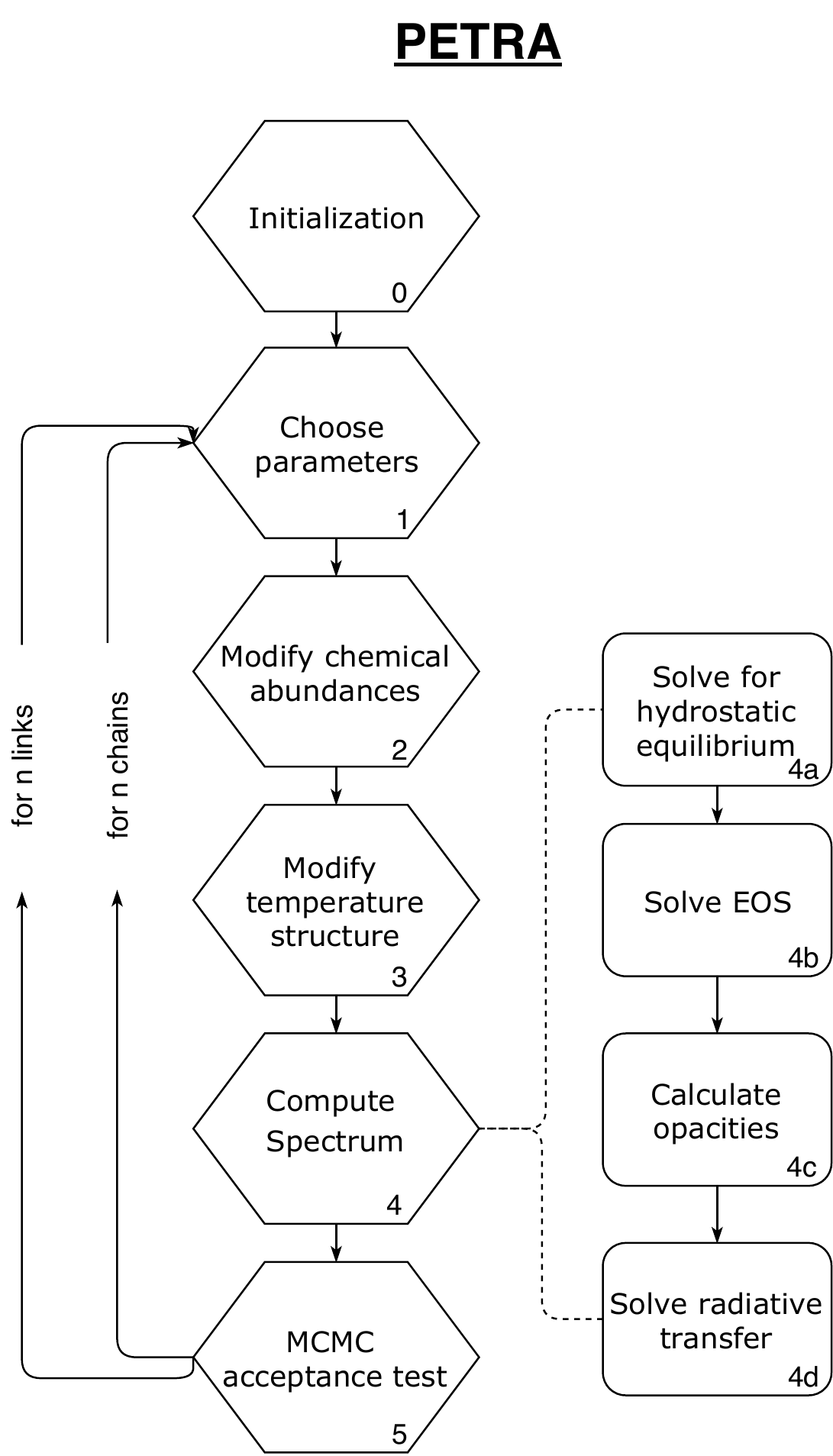}
	\caption{A flow chart illustrating the structure of PETRA. Boxes 4a-d represent the steps that directly utilize PHOENIX. \label{petra_flow}}
\end{figure}

\section{Methods}\label{sec:methods}

\subsection{The Forward Model}

As mentioned above, the forward model uses a modified version of the atmosphere code PHOENIX. In its widely-used self-consistent version, PHOENIX is capable of modeling a variety of objects from cool, cloudy brown-dwarfs to dwarf and giant stars in both plane-parallel and spherically symmetric geometry. This breadth in applicability extends to PETRA as well. Figure~\ref{petra_flow} illustrates the basic structure of PETRA. The left-hand column represents the statistical framework, while the right-hand column shows the steps that utilize PHOENIX for calculating the spectrum.

One of the main motivations for developing PETRA is to utilize PHOENIX's expansive opacity database and opacity sampling routines. PHOENIX has molecular opacity information including many isotopologues, deuterated species, and multiple line lists. Comprehensive atomic line lists are used for elements from hydrogen to uranium, including many bound-free and continuous opacities. More information on the opacities used that are relevant to exoplanet modelling can be found in \cite{lothringer:2018b} and \cite{lothringer:2019}. \added{New line lists are currently being incorporated into PHOENIX and PETRA, including HITEMP H$_2$O and CO$_2$ \citep{rothman:2010}, AlO \citep{patrascu:2015}, TiO \citep{mckemmish:2019}, and SH \citep{zahnle:2009b,yurchenko:2018}}.

PHOENIX uses direct opacity sampling (dOS) to calculate opacities. dOS operates by taking line list information and selecting all relevant lines above some flexible cutoff in opacity relative to the continuum. This is done at the beginning of each model run without the need for precomputed opacity tables, allowing a high degree of flexibility when it comes to changing model resolution, line lists, and included opacity sources. This flexibility will allow PETRA to retrieve quantities from high-resolution observations at high accuracy. For each iteration, the selected lines are broadened and the opacity is then sampled on a user-defined wavelength grid.  

Within PHOENIX, line profiles can be calculated as Gaussian, Voigt, or special line profiles depending on the line's strength. In all cases, the lines are broadened accounting for the natural line profile, quadratic Stark damping, Van der Waals damping, and thermal broadening. The linear Stark effect on atomic hydrogen lines is accounted for by using special line profiles. The Voigt profiles can be described as 

\begin{equation}
V(u,a) = \frac{(\lambda^{2}/c)}{\pi^{1/2}\Delta\lambda_D} H(u,a)
\end{equation}

\noindent with $u$ = $\Delta\lambda$/$\Delta\lambda_D$ and $a$ = $(\lambda^{2}\gamma/4\pi{c})/\Delta\lambda_D$, where $\lambda$ is the wavelength, $\Delta\lambda$ is the distance from the line center, $\Delta\lambda_D$ is the thermal broadening Gaussian width, and $c$ is the speed of light \citep{gray:1992,schweitzer:1996}. $\gamma$ is the Lorentzian damping constant, which is the sum of the natural, quadratic Stark, and Van der Waals broadening widths. Lastly, $H(u,a)$ is the Hjerting function \citep{hjerting:1938}, given by 

\begin{equation}
H(u,a) = \frac{a}{\pi}\int_{-\infty}^{\infty} \frac{e^{-u'^2}}{(u-u')^2+a^2} du'.
\end{equation}

The calculation of the Voigt profile can be computationally expensive when done for many thousands of lines on the fly, so Gaussian profiles are sometimes assumed\added{ for weak lines}. In Section~\ref{sec:wasp43previous}, we test whether Gaussian profiles can be assumed within PETRA to speed up the calculation of the spectrum. Even with calculating Voigt profiles, PETRA can complete a single iteration modeling the CO bandhead from 2.3-2.7 \microns{} at 0.1 \AA\xspace sampling (R$\sim$230,000,$\sim 0.01$cm$^{-1}$) in about 2 seconds on a single 28-core node of a high-performance computer.

Chemical equilibrium is calculated with PHOENIX's Astrophysical Chemical Equilibrium Solver (ACES) which we use to calculate the equation of state for 894 different species. In the PETRA retrievals shown below, the chemical abundances are read from a precomputed partial pressure table and then modified for the species treated as free parameters. PETRA is also capable of retrieving abundances self-consistently, using quantities like the metallicity, elemental, or even isotopic ratios as free parameters.

\subsubsection{Temperature Structure}\label{sec:tempstruct}

In the retrieval of Earth, solar system, and brown dwarf atmospheric observations, there is generally enough data for the temperature structure to be determined layer-by-layer \citep[e.g.,][]{gottwald:2011,irwin:2008,line:2014b}. The low signal-to-noise nature of exoplanet atmosphere observations often necessitates reducing the number of free parameters as much as possible. It is therefore advantageous to parameterize the temperature structure.

In PETRA, we have incorporated n-layer models (where each layer is connected via a logarithmic temperature gradient), the parameterization of \cite{madhusudhan:2009}, and the parameterization used in \cite{line:2013}. The latter is an analytic parametrization for atmospheres in radiative equilibrium using a three-channel Eddington approximation from \cite{parmentier:2014}. Because of the physical motivation of this parameterization and its flexibility, we adopt it for this work.

\subsubsection{Non-Uniform Vertical Abundances}\label{sec:nonuni}

In exoplanet atmosphere retrievals, vertical abundances are often assumed to be constant with pressure. This assumption can break down at high temperatures in ultra-hot Jupiters due to the thermal dissociation of molecules and at temperatures near the transition in chemical equilibrium between CH$_4$ and CO as the dominant carbon-bearing molecule. Within PETRA, the vertical abundances can be described by three parameters: $\eta _{\rm{max}}$ (the maximum volume mixing ratio (VMR) of the species), $\epsilon$ (the power of the slope), and $\eta_{0}$ (the abundance at log(P$_{cgs}$)=0 (i.e., $P=1$ $\mu$bar = 1 barye)). The VMR is thus \deleted{be }parameterized as:
\begin{equation}
\rm{log}_{10}(\rm{VMR}) =min\left(\eta _{\rm{max}},\epsilon*log_{10}(P)+\eta_{0}\right).
\end{equation}
For example, in a typical ultra-hot Jupiter undergoing molecular dissociation at pressures below 1 mbar, CO would have $\eta _{\rm{max}} \sim{-3.5}$, $\eta_{0} \sim{-7}$, and $\epsilon$ would be positive (the molecular abundance would be increasing with pressure).
For species like ions that generally decrease with pressure at photospheric depths, $\epsilon$ would be negative.

This parameterization is clearly limited as there is no physical reason to assume a power-law slope. Future studies may explore layer-by-layer abundance retrievals. This parameterization, however, provides necessary flexibility and a fundamental insight into whether atmospheric species are increasing or decrease with pressure, which will be important in Section~\ref{sec:hminus}. 

\subsection{MCMC/Statistical Framework}

We utilize Differential Evolution Markov Chains (DEMC) \citep{terbraak:2006} with ``snooker" updates \citep{terbraak:2008} to explore the parameter space and build our posterior distributions. DEMC is a type of Markov Chain Monte Carlo (MCMC) that uses information from other chains to determine the next step's size and direction, which is an improvement on the random direction and manual step size of traditional Metropolis-Hastings MCMC. DEMC has been tested and used for a number of exoplanet retrieval applications \citep{line:2013,evans:2017}. 

We have also implemented a version of parallel tempering in PETRA. At the beginning of a retrieval when comparing the likelihoods of the current and proposed state, we raise each likelihood to a power, $\zeta$, such that it is accepted with probability

\begin{equation}
p = min\left(1,\frac{\pi(\boldsymbol{\theta}^{\prime})\mathcal{L(\boldsymbol{\theta}^{\prime}})^{\zeta}}{\pi(\boldsymbol{\theta})\mathcal{L(\boldsymbol{\theta}})^{\zeta}} \right) ,
\end{equation}

\noindent where $\pi(\boldsymbol{\theta}^{\prime})$ is the prior on the proposed state, $\boldsymbol{\theta}^{\prime}$, $\pi(\boldsymbol{\theta})$ is the prior on the current state, $\boldsymbol{\theta}$, $\mathcal{L(\boldsymbol{\theta}^{\prime}})$ is the likelihood of the proposed state, $\mathcal{L(\boldsymbol{\theta}})$ is the likelihood of the current state, $\boldsymbol{\theta}$, and $\zeta$ is the `temperature', which we define from 0 to 1. $\zeta=1$ corresponds to no tempering while $0<\zeta<1$ ``flattens" posterior space. Flatter posterior distributions help the chains explore different local likelihood minima regions without getting stuck. We generally begin each retrieval with a short period of tempering where $\zeta$ begins at a minimum, $\zeta_{min}$, usually between 0.5 and 0.9 and slowly rises for $n_{max}$ iterations  until $\zeta=1$:

\begin{equation}
\zeta = (\zeta_{min}-1)*\left(1-\frac{n}{n_{\rm{max}}}\right)+1,
\end{equation}

\noindent where $n$ is the current iteration. We usually set $n_{max}$ between 100 to 1,000 so tempering only occurs during the initial burn-in of the chains. Our choice of $n_{max}$ helps ensure that we find the global likelihood maximum.

In a full retrieval using the temperature structure parameterization described above and four to eight molecular abundances as free parameters for a total of 9 to 13 free parameters, we use about 20 chains and can reach convergence, as indicated by a Gelman-Rubin statistic $<1.01$ \citep{gelman:1992}\footnote{In the simplest terms, the Gelman-Rubin statistic compares the standard deviation of each chain with the standard deviation of the mean of the chains.}, in $<$$10^4$ iterations per chain.

In future work, we will incorporate nested sampling, which is a powerful method to do Bayesian model comparison \citep{skilling:2004}. This is useful to determine if the complexity of the model\added{, i.e., the number of free parameters,} is justified.

\section{Tests}\label{sec:tests}

\subsection{Simulated Data from PHOENIX}

To validate the statistical framework of PETRA, we used our forward model to compute a hot Jupiter spectrum with parameters similar to those of WASP-43b.  From this spectrum, we simulated observations from the \textit{Hubble Space Telescope}'s Wide Field Camera 3 (HST/WFC3) and \textit{Spitzer} Channels 1 and 2 using the bins and uncertainties from \cite{kreidberg:2015}. We then retrieved the atmospheric parameters using PETRA from these simulated data for which we know the `true' values. Our retrieval uses the same parameters as the input model: five parameters to describe the temperature profile (discussed above) and four parameters for the H$_2$O, CO, CO$_2$, and CH$_4$ abundances.

\begin{figure}
	\center    
	\includegraphics[width=3.4in]{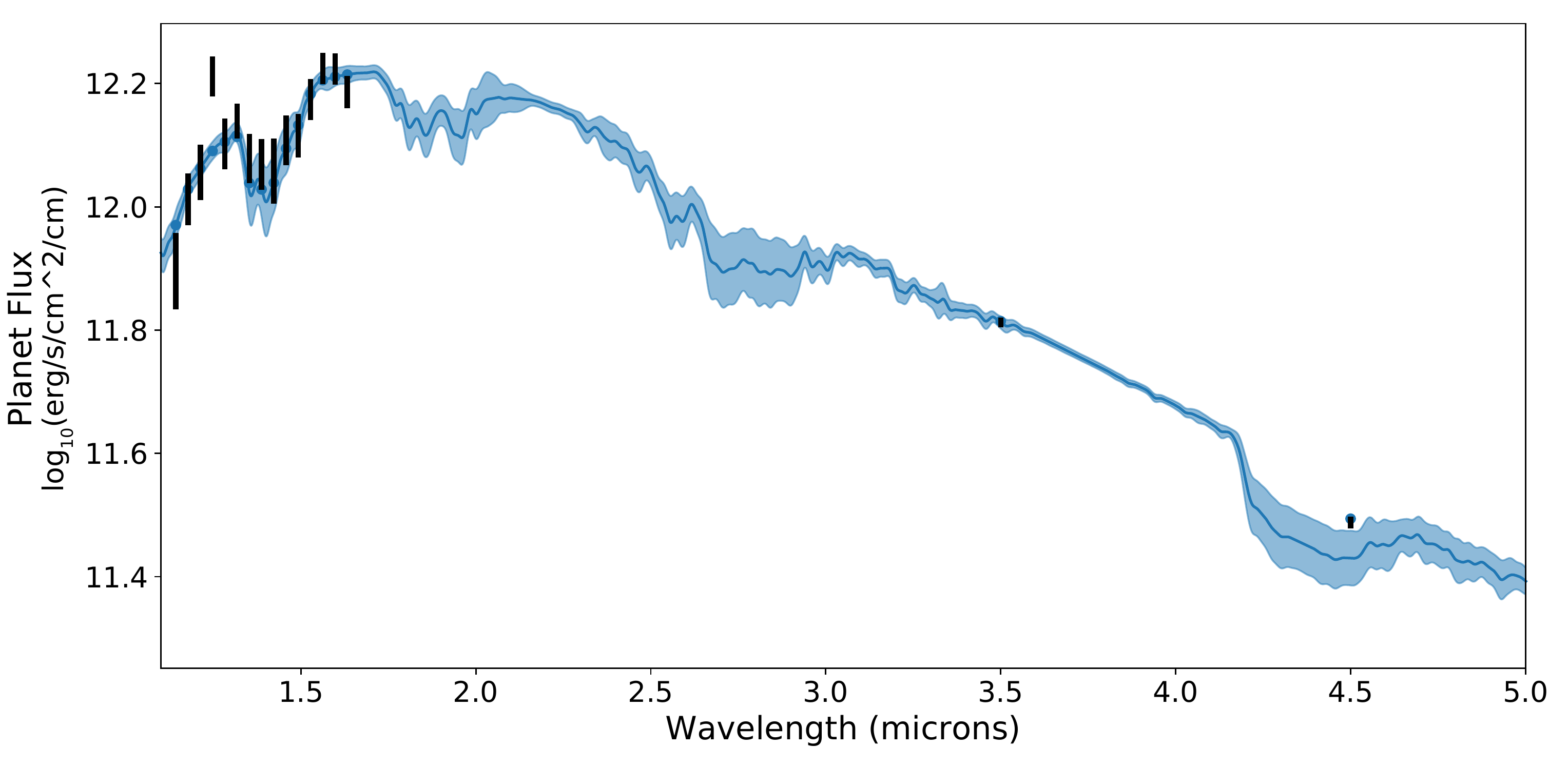}
	\includegraphics[width=3.4in]{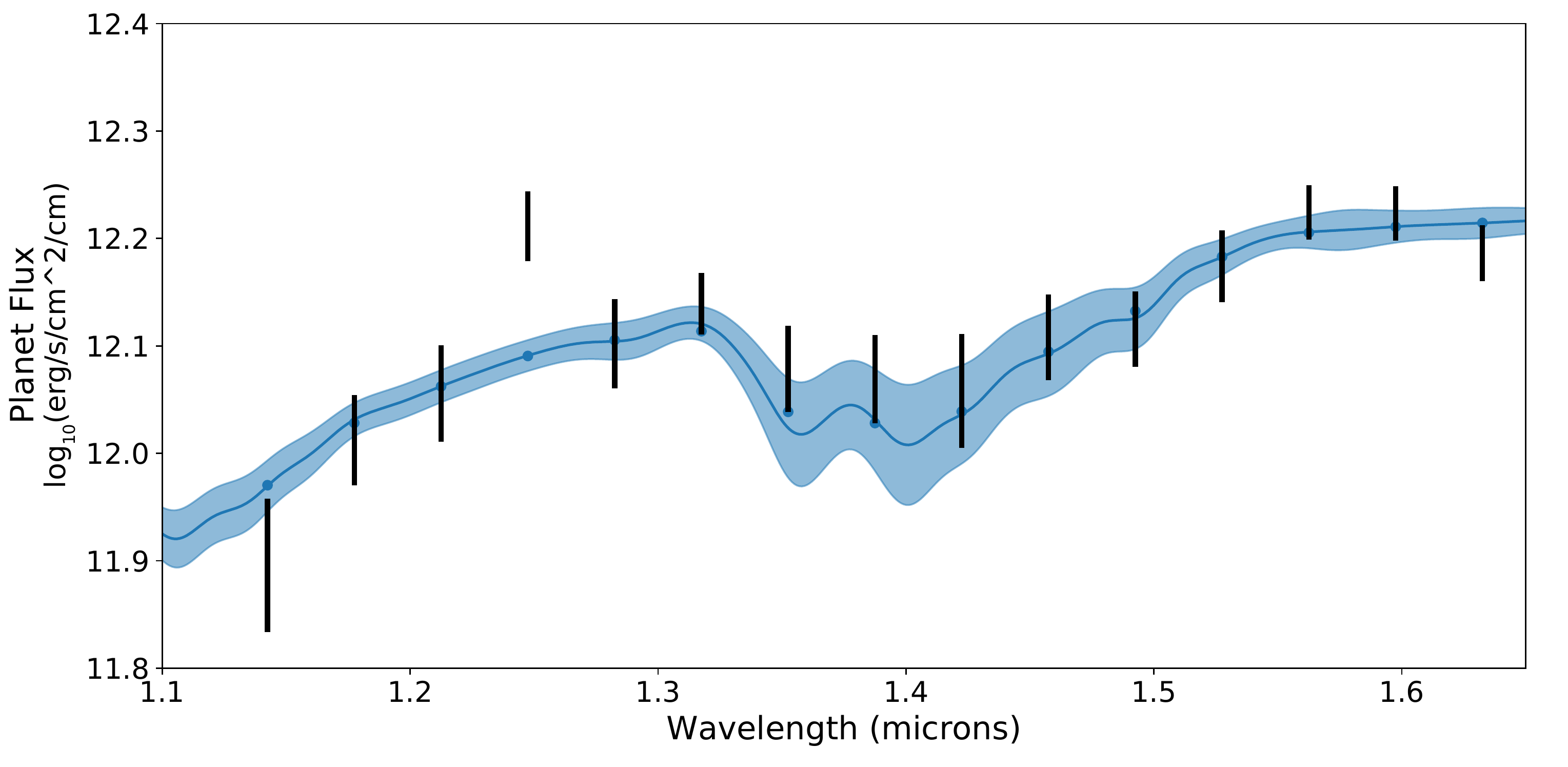}
	\caption{Simulated observations of a generic hot Jupiter similar to WASP-43b from a PHOENIX forward model compared to the retrieved median spectrum and 1-$\sigma$ uncertainty region. The bottom figure is a zoomed in version of the top, highlighting the 1.4 micron H$_2$O feature from HST/WFC3 observations. The region between the end of the HST/WFC3 bandpass ($\sim 1.7$ \microns{}) and the beginning of the \textit{Spitzer} Channel 1 bandpass ($\sim 3$ \microns{}) was sampled sparsely for computational efficiency. \label{fig:w43spec}}
\end{figure}

Figure~\ref{fig:w43spec} shows the simulated observations compared to the median retrieved spectrum and the spectrum's 1-$\sigma$ uncertainty region. PETRA is able to fit the simulated data well, with \chisq{} = 17.23 for the median retrived spectrum. With 17 data points, we obtain a \chisq{} per data point of 1.01, quite similar to the quality of fits obtained with real data in the sections below.

\begin{figure}
	\center   
	\includegraphics[width=3.5in]{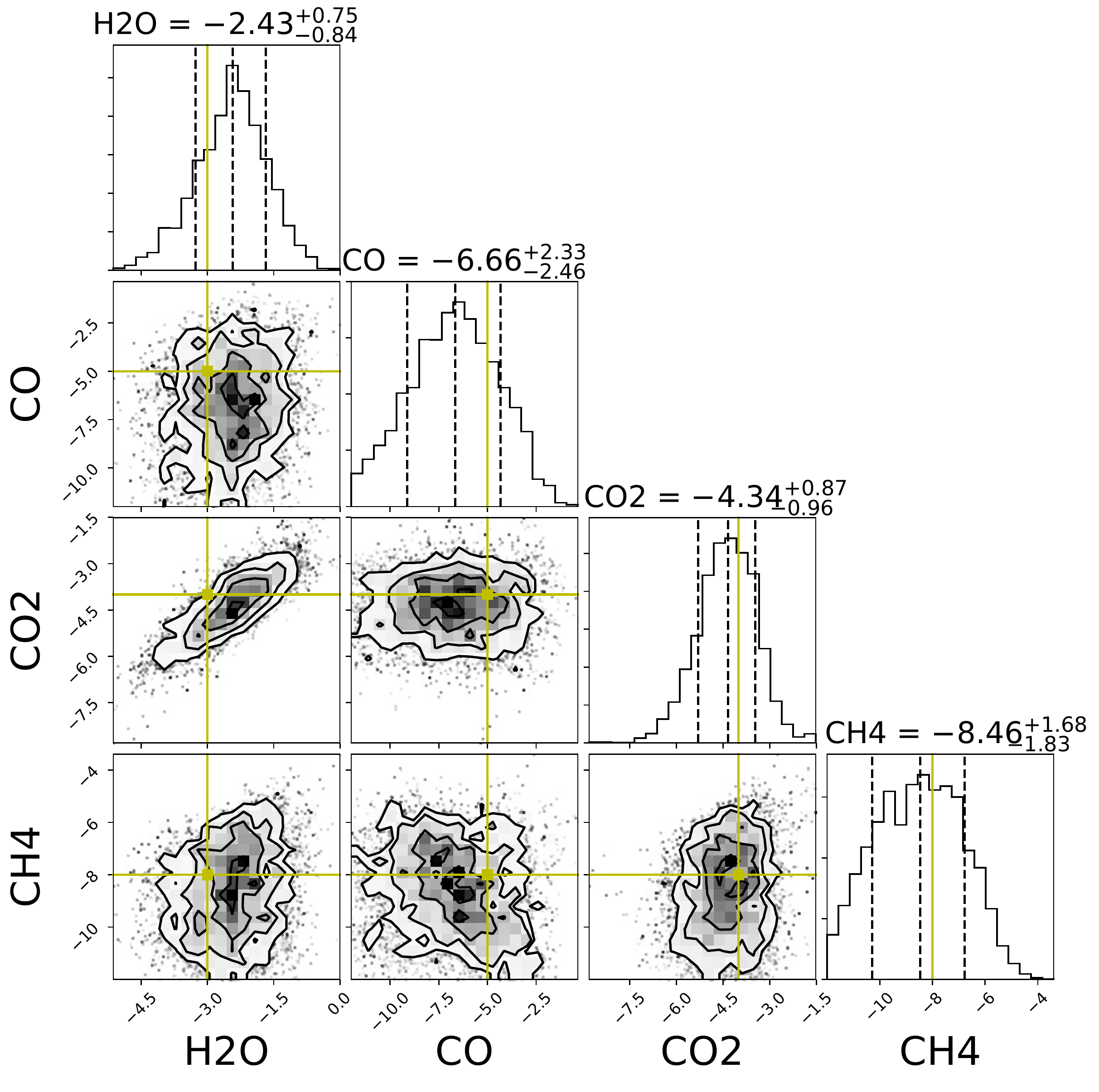}
	\caption{Plot of the posterior distributions of molecular abundances from the WASP-43b simulated data test. The top-most plot of each column shows the 1-D posterior for each molecular abundance included in the test. The middle dotted line shows the average retrieved value, while the dotted lines to each side bound the 1-$\sigma$ uncertainty range. The yellow lines indicate the `true' input value that generated the simulated data. Below the 1-D posteriors are the 2-D posterior distributions.\label{fig:w43mol}}
\end{figure}

Figure~\ref{fig:w43mol} shows the retrieved constraints on the molecular abundances. The top plot in each column shows the 1-dimensional posterior distribution for each molecular abundance. Below the 1-dimensional posteriors are 2-dimensional posteriors showing how the estimates of each molecular abundance depend on the other molecular abundances. The retrieved molecular abundances agree to within 1-$\sigma$ of the input molecular abundance, indicating that we are retrieving correct values.

\begin{figure}
	\center    
	\includegraphics[width=3.5in]{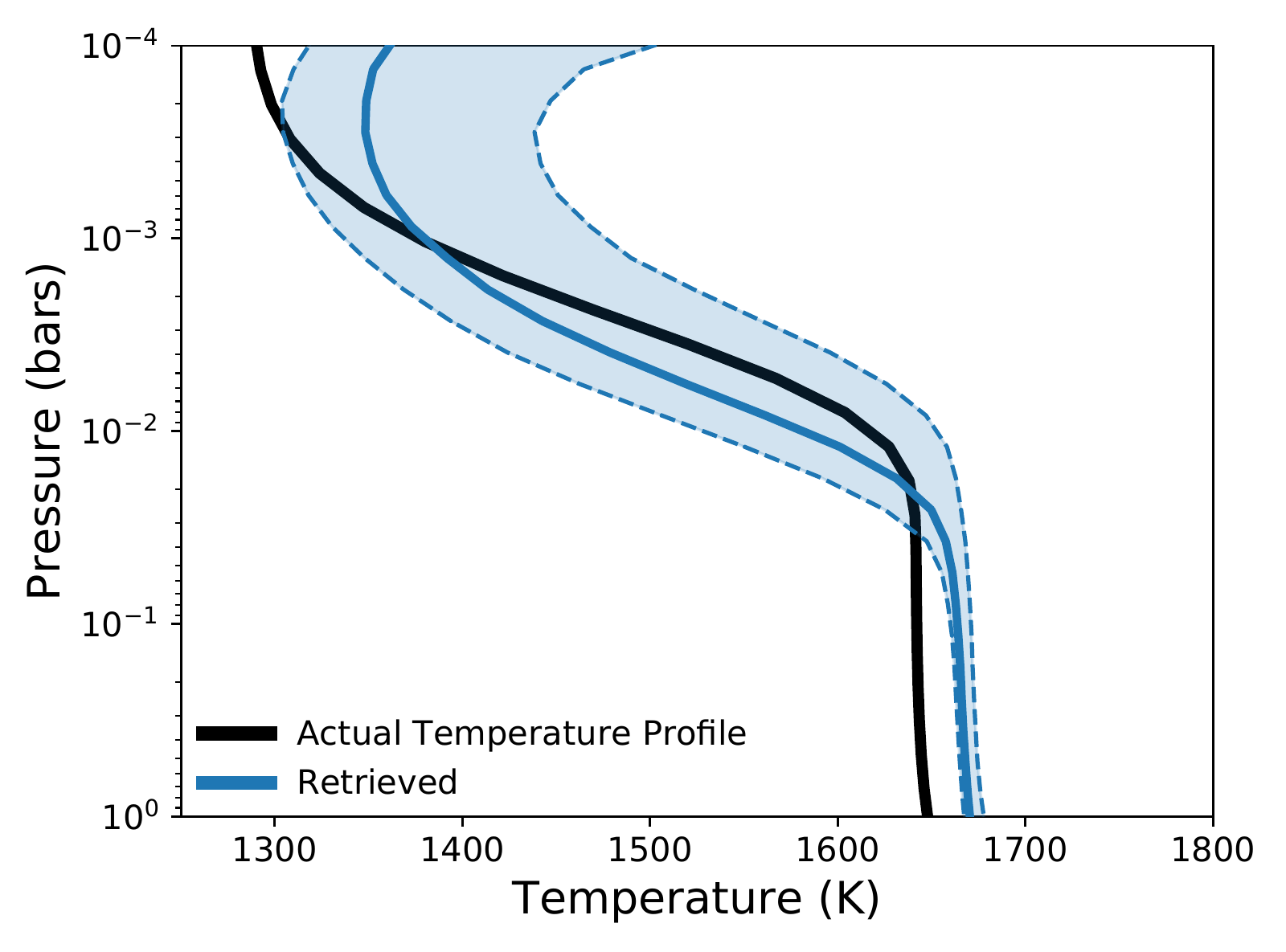}
	\caption{Comparison of the input temperature profile (gold) for the WASP-43b simulated data test compared to the average retrieved temperature profile (dashed black) and the 1-$\sigma$ uncertainty region (shaded grey region). The region above 1~mbar is not probed by the observations and the constraints shown are an outcome of the parameterization.\label{fig:w43tstruct}}
\end{figure}

Figure~\ref{fig:w43tstruct} compares the retrieved temperature structure constraints with the input temperature profile. The retrieved temperature structure is within 1 to 2-$\sigma$ of the input temperature profile in the optical and IR photosphere region between 1-100 mbar. The upper atmosphere, above the optical and IR photosphere deviates from the input temperature profile, but since none of the simulated observation probe this region, this is acceptable. Retrieved information about atmospheric depths not probed by the observations is often biased by the parameterization of the temperature profile and, thus, do not provide reliable constraints.

\begin{figure}
	\center  
	\includegraphics[width=3.0in]{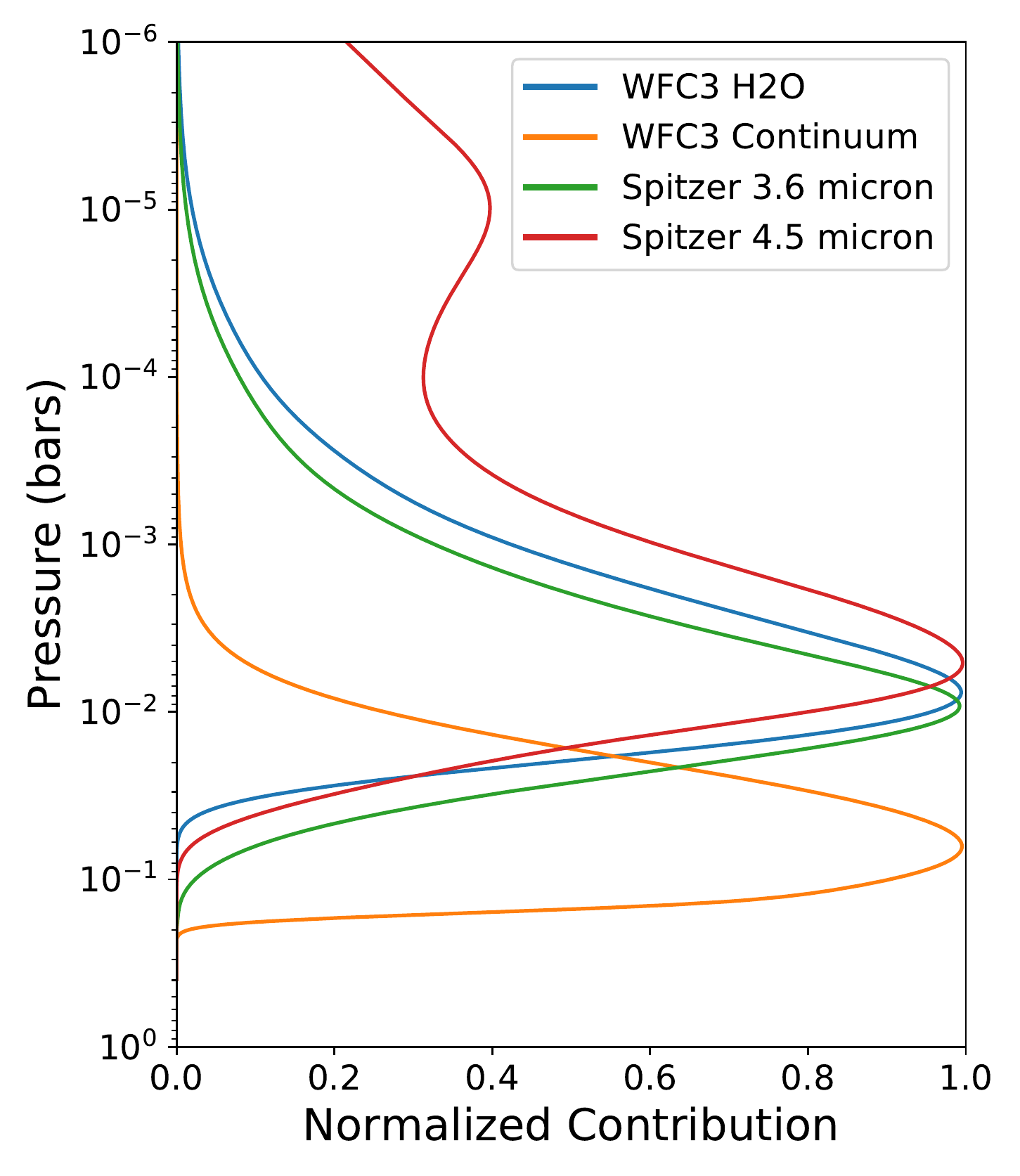}
	\caption{Contribution functions for WASP-43b's dayside atmosphere at 1.25, 1.4, 3.6, and 4.5 \microns{}, corresponding to outside the 1.4 \microns{} H$_2$O feature, inside the 1.4 \microns{} H$_2$O feature, \spitzer{} channel 1, and \spitzer{} channel 2, respectively. \label{fig:w43_cf}}
\end{figure}

The region being probed by the observations can also be understood by looking at the contribution functions of the atmosphere at wavelengths corresponding to the observations. Figure~\ref{fig:w43_cf} shows contribution functions at HST/WFC3 wavelengths inside and outside the H$_2$O absorption feature, as well as the two \spitzer{} photometry points. The peak of each contribution function shows from which pressure most of the flux at a given wavelength is radiated. While the HST/WFC3 continuum probes deep in the atmosphere at around 100 mbar, the other wavelengths probe higher up the atmosphere (due to increased opacity at these wavelengths) at around 5 mbar. Therefore, the WASP-43b's temperature is only constrained by the observations between about 1 and 100 mbar.

\subsection{Comparison to Previous Retrievals}

\subsubsection{WASP-43b}\label{sec:wasp43previous}

\begin{figure}
	\center    
	\includegraphics[width=3.5in]{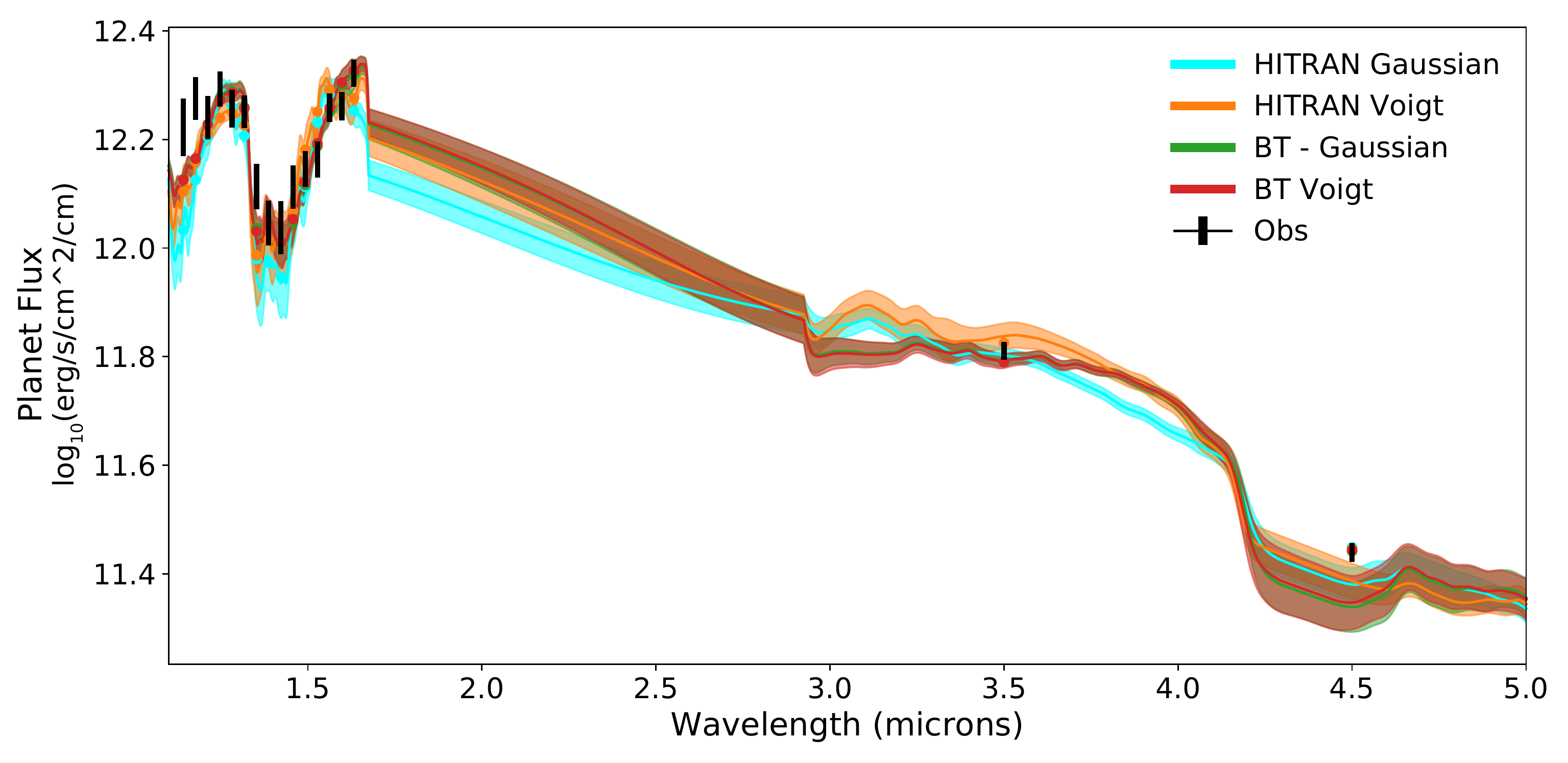}
	\includegraphics[width=3.5in]{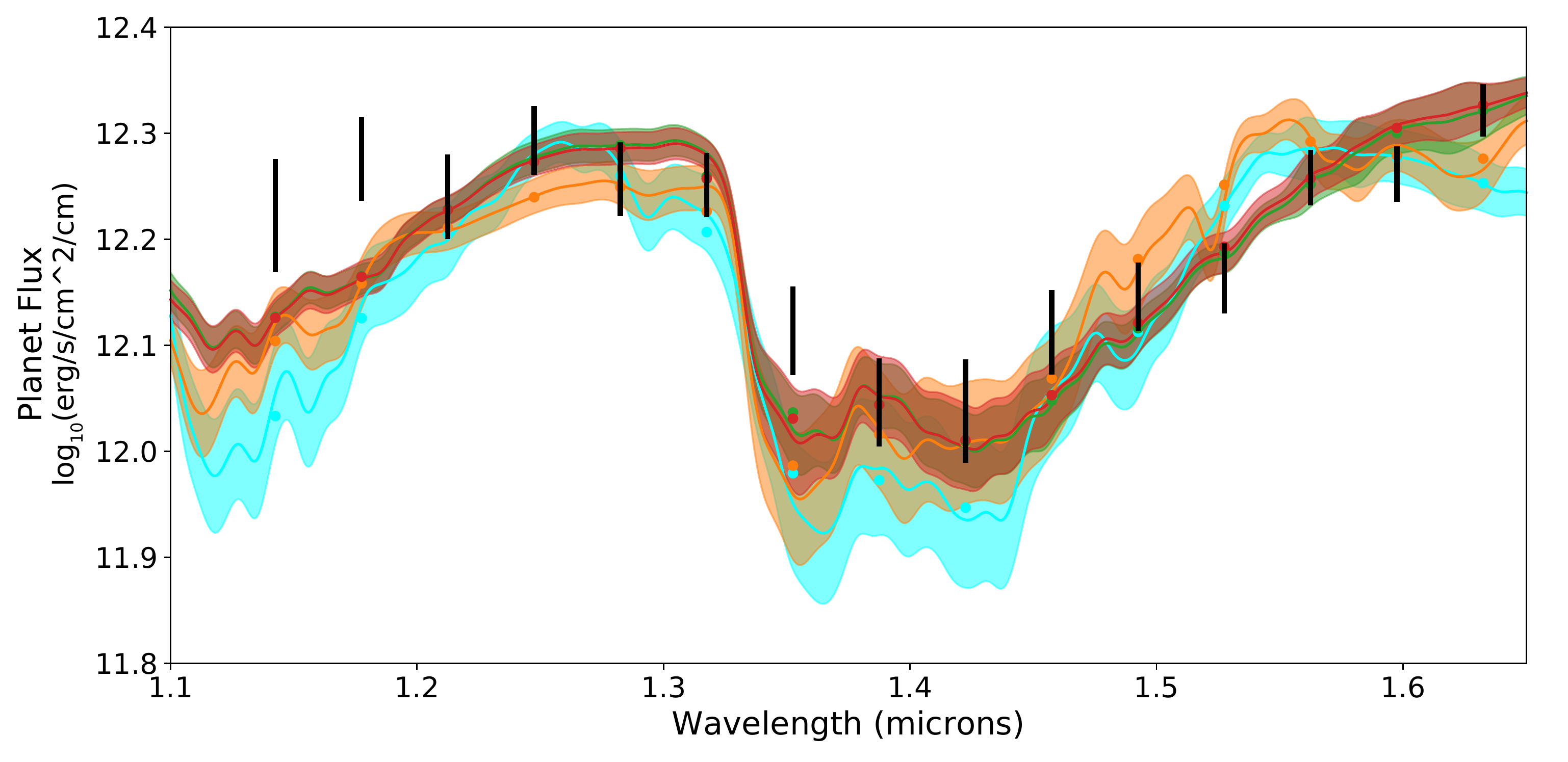}
	\caption{Observations of hot Jupiter WASP-43b from \cite{kreidberg:2015} compared to the retrieved median spectrum and 1-$\sigma$ uncertainty region for retrievals using different H$_2$O line lists and line profile treatment (see text).The region between the end of the HST/WFC3 bandpass ($\sim 1.7$ \microns{}) and the beginning of the \textit{Spitzer} Channel 1 bandpass ($\sim 3$ \microns{}) was sampled sparsely for computational efficiency. \label{fig:w43realspec}}
\end{figure}

\begin{figure*}
	\center    
	\includegraphics[width=3.5in]{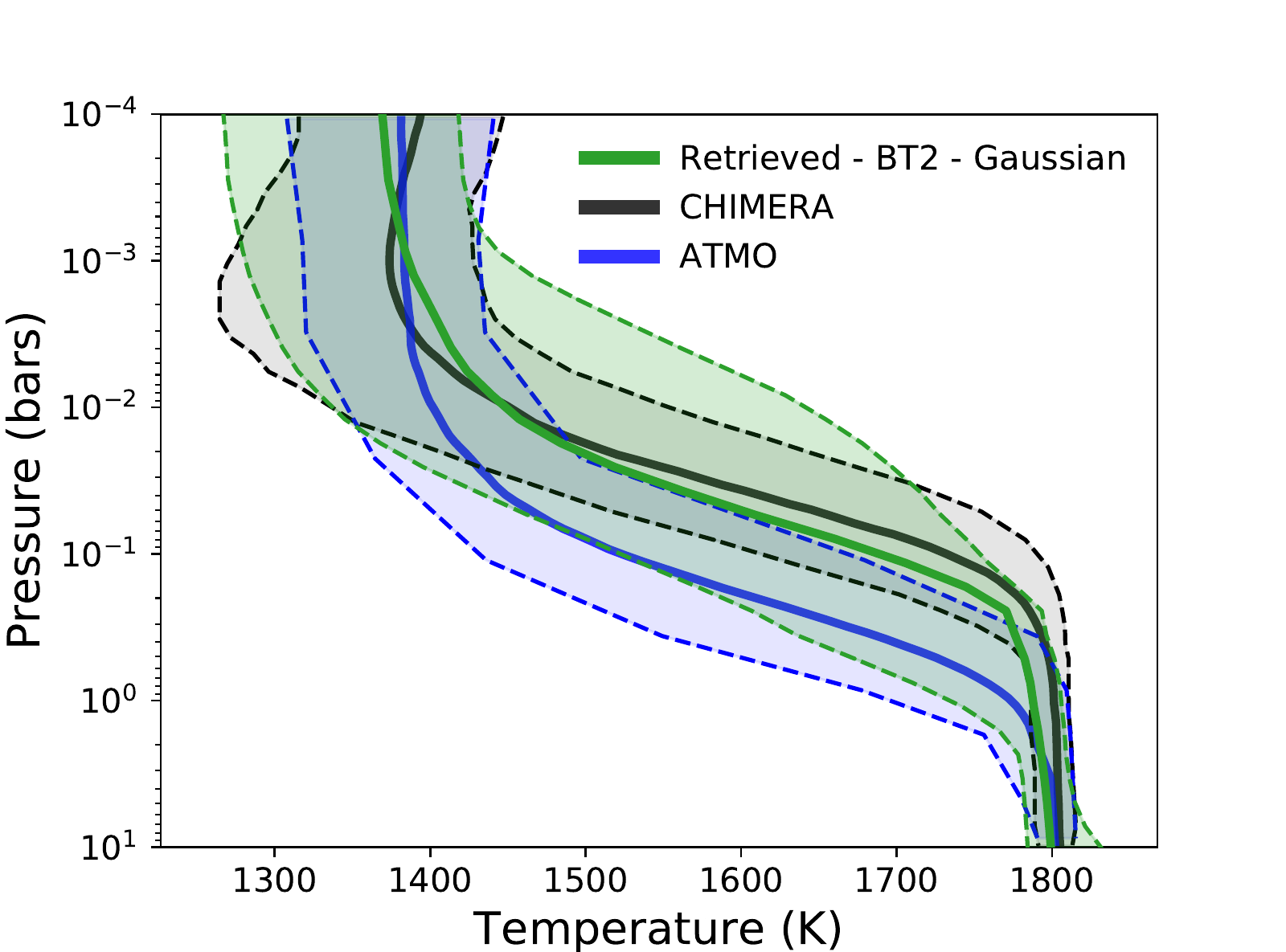}
	\includegraphics[width=3.5in]{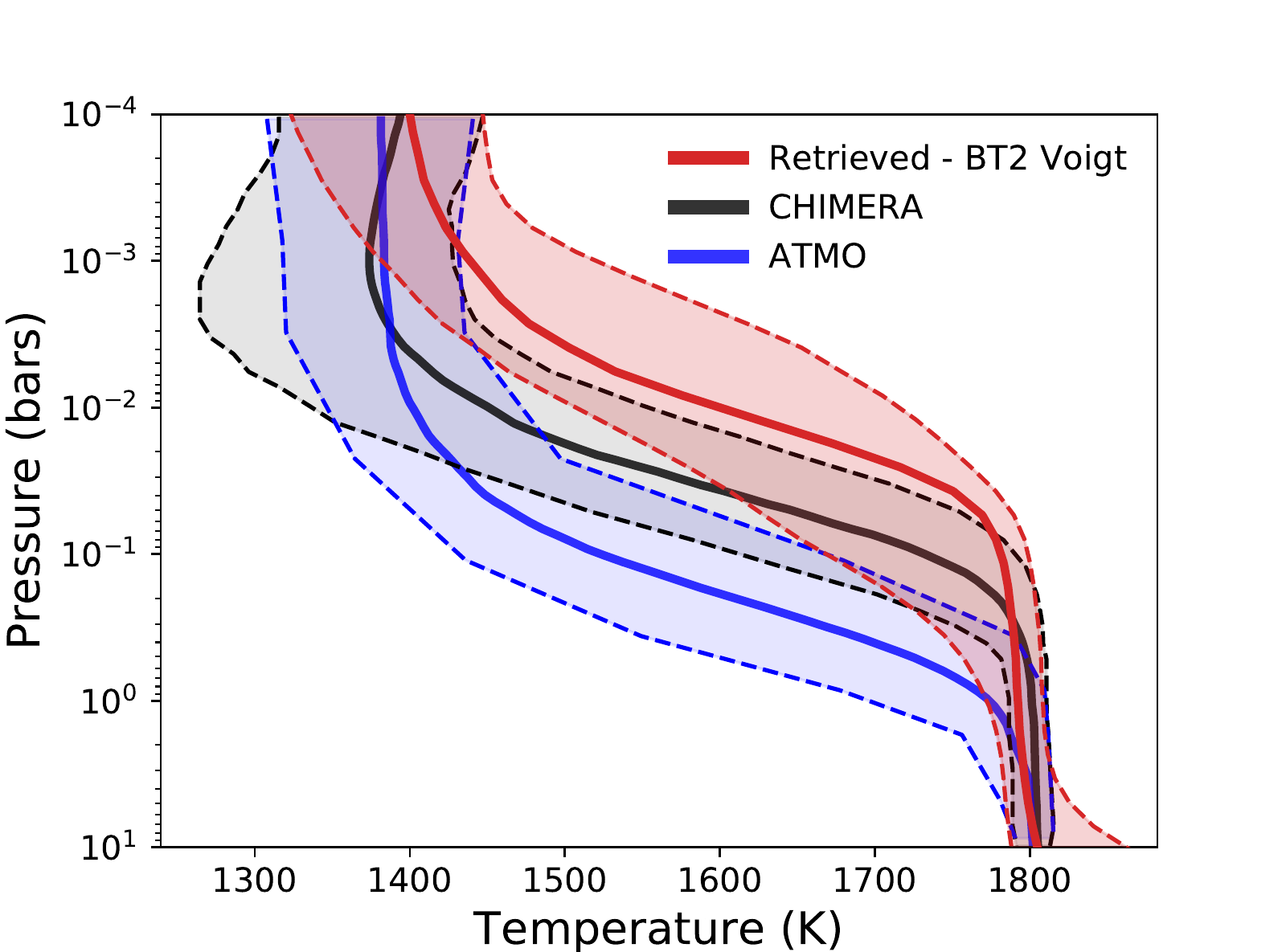}
	\includegraphics[width=3.5in]{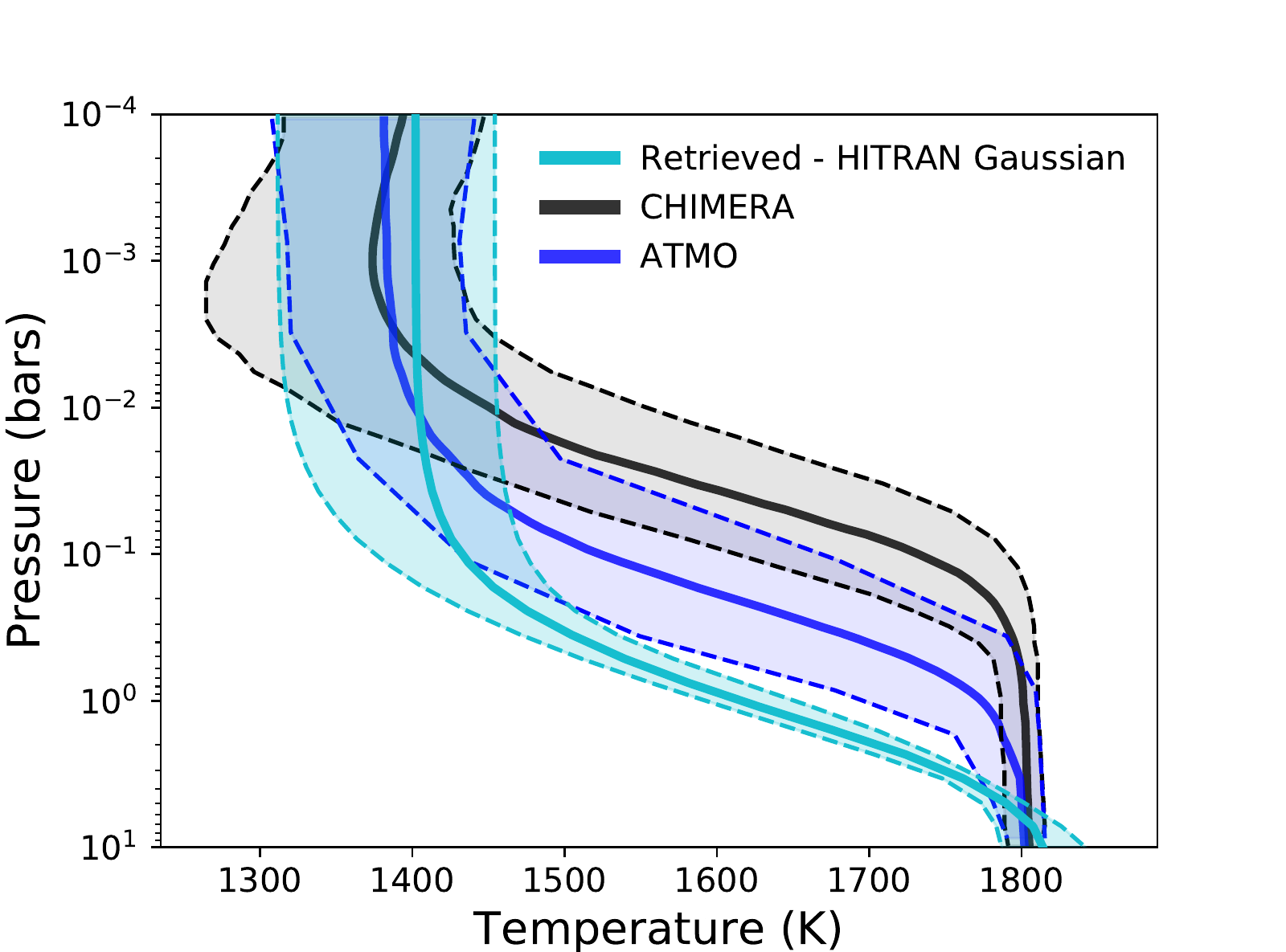}
	\includegraphics[width=3.5in]{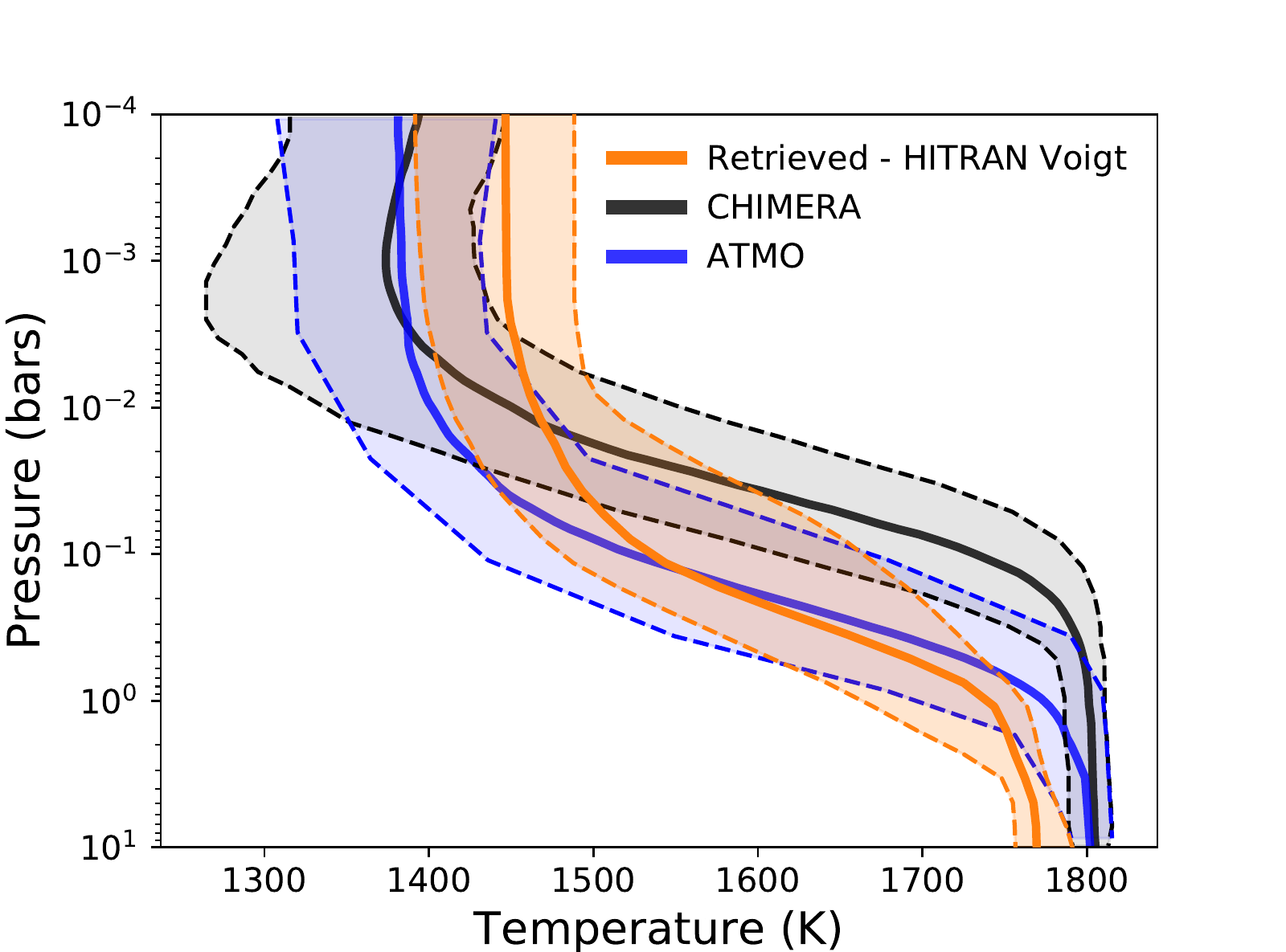}
	\caption{Comparison of the retrieved temperature profiles from real WASP-43b HST/WFC3 and Spitzer data from \cite{kreidberg:2014b} using different H$_2$O line lists and line profile treatment (see text) compared to previous median retrieved temperature profiles (CHIMERA and ATMO from \cite{kreidberg:2014b} and \cite{evans:2017}, respectively). The colors for the PETRA retrievals match those in Figure~\ref{fig:w43realspec}. \label{fig:w43realtstruct}}
\end{figure*}

\begin{figure}
	\center  
	\includegraphics[width=3.4in]{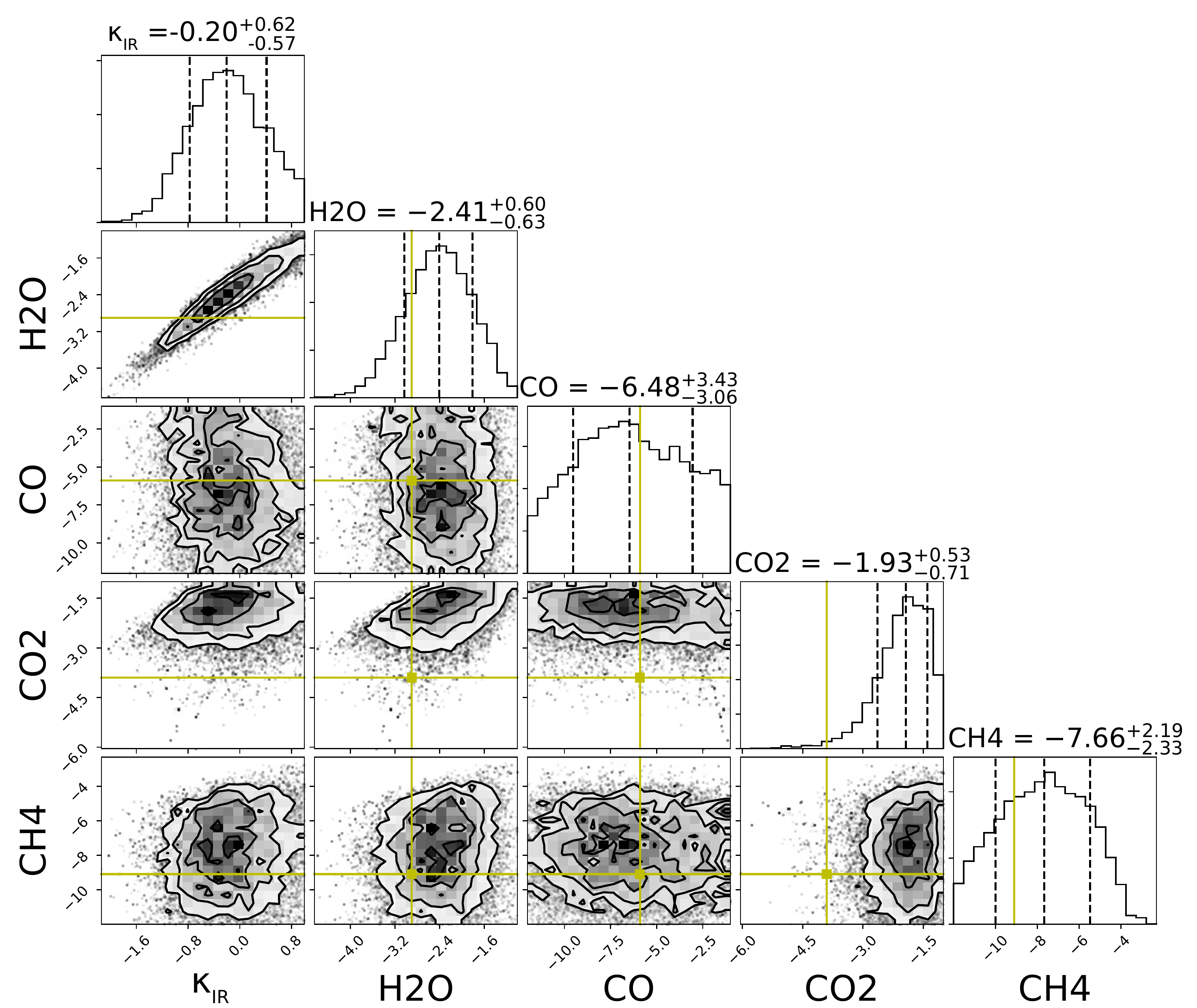}
	\caption{\added{Posterior distribution for some parameters in the WASP-43b BT2 Voigt profile retrieval. Gold lines indicate the values found in \cite{kreidberg:2014b}.} \label{fig:w43_real_post}}
\end{figure}

\begin{figure*}
	\center    
	\includegraphics[width=3in]{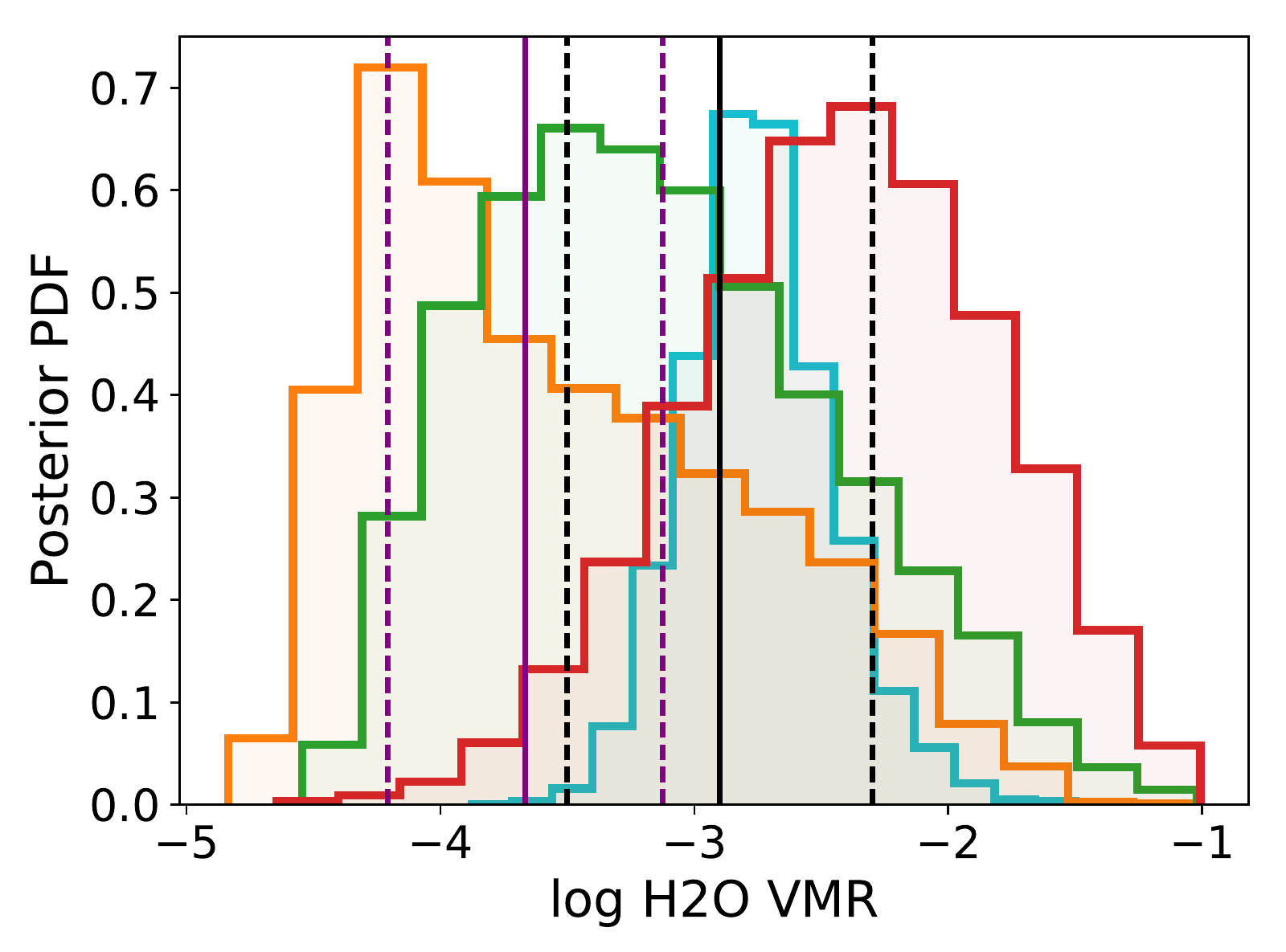}
	\includegraphics[width=3in]{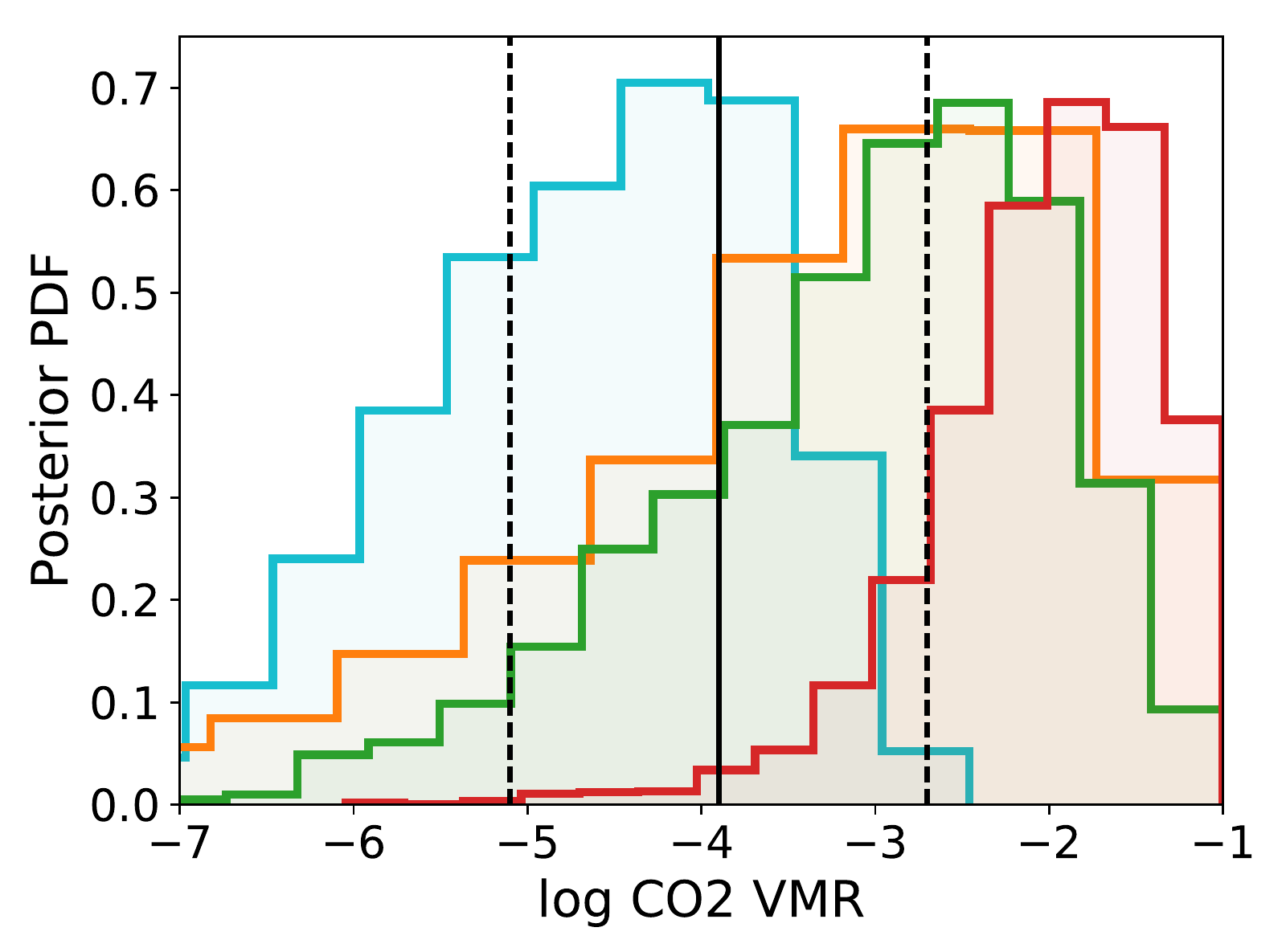}
	\caption{Posterior distributions of H$_2$O and CO$_2$ from real WASP-43b HST/WFC3 and \textit{Spitzer} data from \cite{kreidberg:2014b} using different H$_2$O line lists and/or molecular profiles, compared to previous retrieved (same colors and labels as Figures~\ref{fig:w43realspec} and \ref{fig:w43realtstruct}). Retrieved abundances and 1-$\sigma$ from \cite{kreidberg:2014b} are in black and from \cite{evans:2017} are in purple. Note that \cite{evans:2017} only presents the H$_2$O abundance.\label{fig:w43realmol}}
\end{figure*}

In order to compare PETRA with other retrieval suites, we retrieved the temperature profile and molecular abundances of hot Jupiter WASP-43b using HST/WFC3 and \textit{Spitzer} data from \cite{kreidberg:2014b}, which was analyzed with the CHIMERA retrieval suite \citep{line:2013,line:2013b}. This same data was also analyzed in \citep{evans:2017} with the ATMO retrieval suite. Both frameworks retrieved a non-inverted atmosphere with an approximately solar abundance of H$_2$O.

As was done in the CHIMERA retrievals, we use the temperature profile parameterization described above and fit for the H$_2$O, CO, CO$_2$, and CH$_4$ abundances. We analyzed the WASP-43b data with PETRA using two different H$_2$O line lists, BT2 (Barber-Tennyson): \cite{barber:2006} and HITRAN2008: \cite{rothman:2008}. BT2 is a computed list of H$_2$O transition frequencies and intensities and comprises over 5$\times 10^6$ transitions. HITRAN2008 is a compilation of transitions from various sources, including over 6.9$\times 10^4$ H$_2$O transitions.

As mentioned in Section~\ref{sec:methods}, the calculation of the Voigt profile can be computationally expensive when done on the fly as in PETRA. \added{Additionally, \citep{barstow:2020} showed that assumptions in line broadening can lead to significant differences in forward models and retrieval results.} \replaced{To test whether Gaussian profiles can provide physically accurate retrieved atmospheric properties and to understand the overall effect the line profile shape may have on retrieval results}{To understand  the overall effect of the line profile shape on retrieval results and to test whether Gaussian profiles can provide physically accurate retrieved atmospheric properties}, we ran retrievals with each list assuming Gaussian and then Voigt line profiles when calculating the opacities.

Figure~\ref{fig:w43realspec} shows how the observations compare to the median retrieved spectra and 1-$\sigma$ uncertainties for each retrieval. The agreement between the models and observations is quite similar for the two retrievals using the BT2 H$_2$O line list. The median spectrum retrieved using the BT2 list with Voigt line profiles has \chisq{} = 17.39, giving a \chisq{} per data point of \chisq{}/17 = 1.023. The median spectrum retrieved using the BT2 list with the Gaussian line profiles matches the data marginally better with \chisq{} = 15.97 and a \chisq{} per data point of 0.939. Both of retrievals are comparable to the \chisq{} per data point of 1.2 found with CHIMERA in \cite{kreidberg:2015}. 

The \replaced{BT2}{HITRAN2008} list\deleted{, especially assuming Gaussian profiles, seems to} provides a somewhat worse fit\deleted{, particularly for the long wavelength half of the 1.4 \microns{} H$_2$O band}, \added{with \chisq{} = 31.54 and a \chisq{} per data point of 1.86 using Gaussian profiles and} with \chisq{} = 39.16 and a \chisq{} per data point of 2.3\added{ using Voigt profiles}. This may provide some evidence that the BT2 list is more \deleted{accurate and }capable of producing hot Jupiter spectra that better match observations. This is expected since the BT2 list is constructed with special attention paid to the high temperatures found in hot Jupiter atmospheres and considers many more transitions.

Each temperature profile in Figure~\ref{fig:w43realtstruct} is qualitatively similar in shape and temperature range. There are some subtle differences, with the BT2 retrievals having a slightly different lapse rate than the BT2  retrieval with Voigt profiles. The HITRAN2008 retrieval with the Gaussian approximation exhibits a highly constrained lower atmosphere, uncharacteristic of the other profile. Additionally, the temperature profile was uniformly moved to lower pressures in the retrievals with Voigt line profiles compared to the same retrieval with Gaussian line profiles. This underestimation is because a Gaussian line profile will systematically underestimate the opacity for a given line, particularly the line wings. This means that the photosphere of the retrieved model will be deeper in the atmosphere when Gaussian line profiles are assumed. The BT2 retrieval with Gaussian line profiles, however, matched very closely with the CHIMERA retrieval, though they used a HITEMP H$_2$O line list \citep{rothman:2010}, presumably with Voigt profiles. The HITEMP line list\added{, an updated version of the HITRAN2008 list more appropriate for hot atmospheres like we consider here,} uses the BT2 line list as its starting point.

The four different retrieved H$_2$O and CO$_2$ abundances are compared in Figure~\ref{fig:w43realmol}. Only the H$_2$O abundance is published in \cite{evans:2017}. In general, all abundances agree quite well and are close to the solar abundance VMR of H$_2$O of about $10^{-3.6}$. The H$_2$O abundance found in the BT2 retrieval with Voigt profiles was slightly higher than the other retrievals. In general, there is a correlation between the pressure of the temperature profile (parameterized through $\kappa_{IR}$ in \cite{line:2013}) and the molecular abundances. When the temperature profile is moved to lower pressures, the molecular abundances, particularly H$_2$O, must compensate by becoming larger\added{ to keep the same brightness temperature at a given wavelength}. \added{This correlation is shown in Figure~\ref{fig:w43_real_post}, which displays the posteriors of the WASP-43b BT2 Voigt profile retrieval. The correlation between $\kappa_{IR}$ and the chemical abundances likely explains some of the differences between the exact temperature profiles and abundances found in the ATMO, CHIMERA, and PETRA retrievals.} 

Three of the PETRA retrievals show a somewhat higher CO$_2$ abundance, but they agree at about the 1-$\sigma$ level with the CHIMERA value, which used the HITEMP database \citep{rothman:2010}. The HITRAN2008 retrieval with Gaussian profiles showed a significantly lower CO$_2$ abundance, closer to the CHIMERA value.
Additionally, note that most of the information for the CO$_2$ and CO abundances relies on the single \textit{Spitzer} 4.5 micron point. PETRA and CHIMERA both seem to prefer fitting the \textit{Spitzer} data with CO$_2$ rather than CO, however. \added{Removing CO$_2$ does not change the goodness of fit, because CO can compensate; however, removing both CO and CO2 changes the Bayesian Information Criterion (BIC) by about 20, indicating strong evidence for their inclusion in the model. The BIC quantifies whether the complexity of a given model is justified by the data by penalizing the likelihood by a factor proportional to the number of free-parameters \citep{schwarz:1978}.}

\added{We also tested whether there was any evidence of a evidence of a vertically non-uniform H2O abundance H$_2$O abundance by running a retrieval using the parameterization described in Section~\ref{sec:nonuni}. While such a retrieval was able to constrain the H$_2$O abundance consistent with the uniform model, we find that the non-uniform model is not justified given the data, with a $\Delta$BIC of 31.95 in favor of the simpler uniform abundance model. This is in agreement with the theoretical expectation that the H$_2$O abundance should be roughly uniform with pressure throughout much of the observable atmosphere in hot Jupiters in this temperature regime.}

\replaced{On the whole, beyond demonstrating PETRA's ability to match previous retrieval analyses, this exercise illustrates the model uncertainties present in retrieval analyses that emerge as a result of, e.g., line list choices.}{On the whole, assuming Gaussian line profiles can provide results qualitatively similar to Voigt line profiles. However because we do not expect thermal broadening to be the main source of line broadening in most exoplanet retrieval applications, assuming Voigt line profiles will provide the more physically accurate results. This exercise illustrates the model uncertainties present in retrieval analyses that emerge as a result of, e.g., line list choices.} While we can assume that Voigt line profiles will provide a more realistic match to observations than Gaussian line profiles, some choices can be more arbitrary. 
When such cases arise, it is best to evaluate observations using a variety of model assumptions whenever possible, which can begin to quantify our systematic model uncertainties.

\subsubsection{HD 209458b}
\begin{figure}
	\center    
	\includegraphics[width=3.5in]{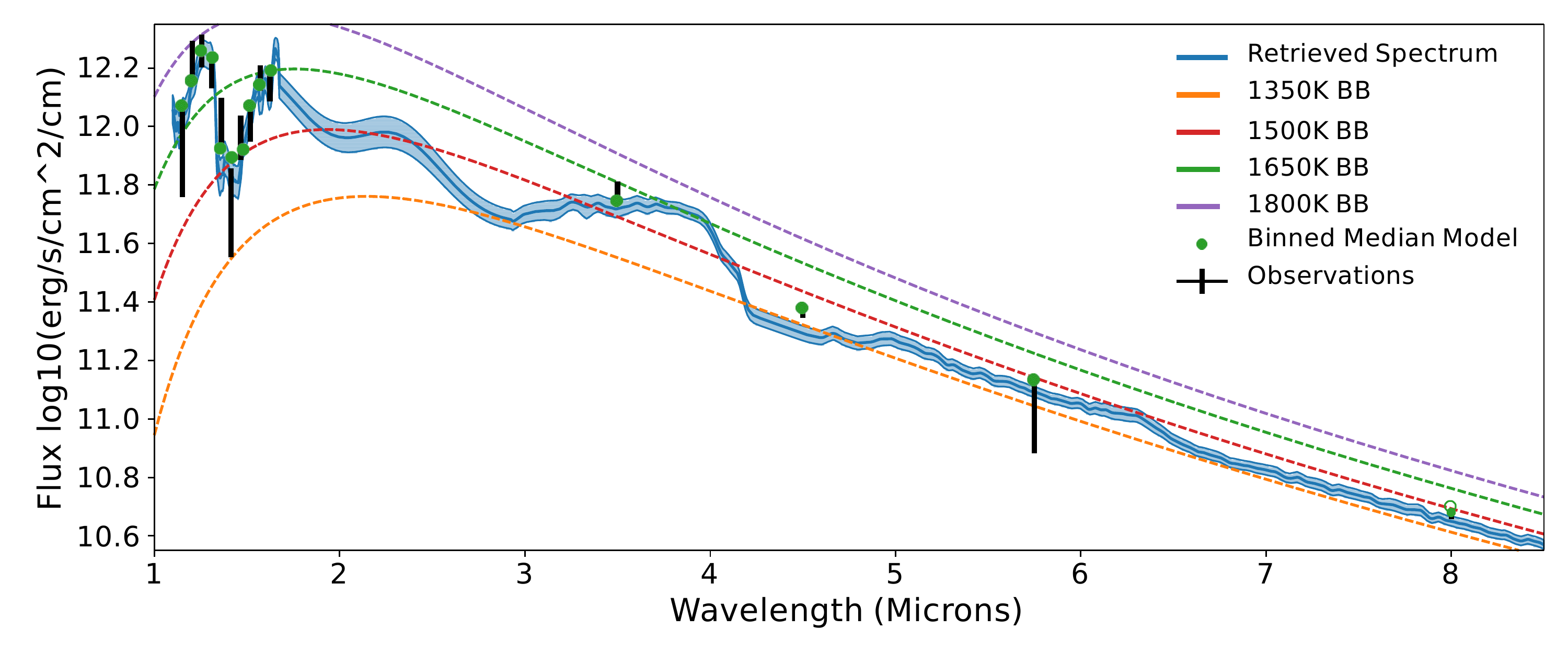}
	\includegraphics[width=3.5in]{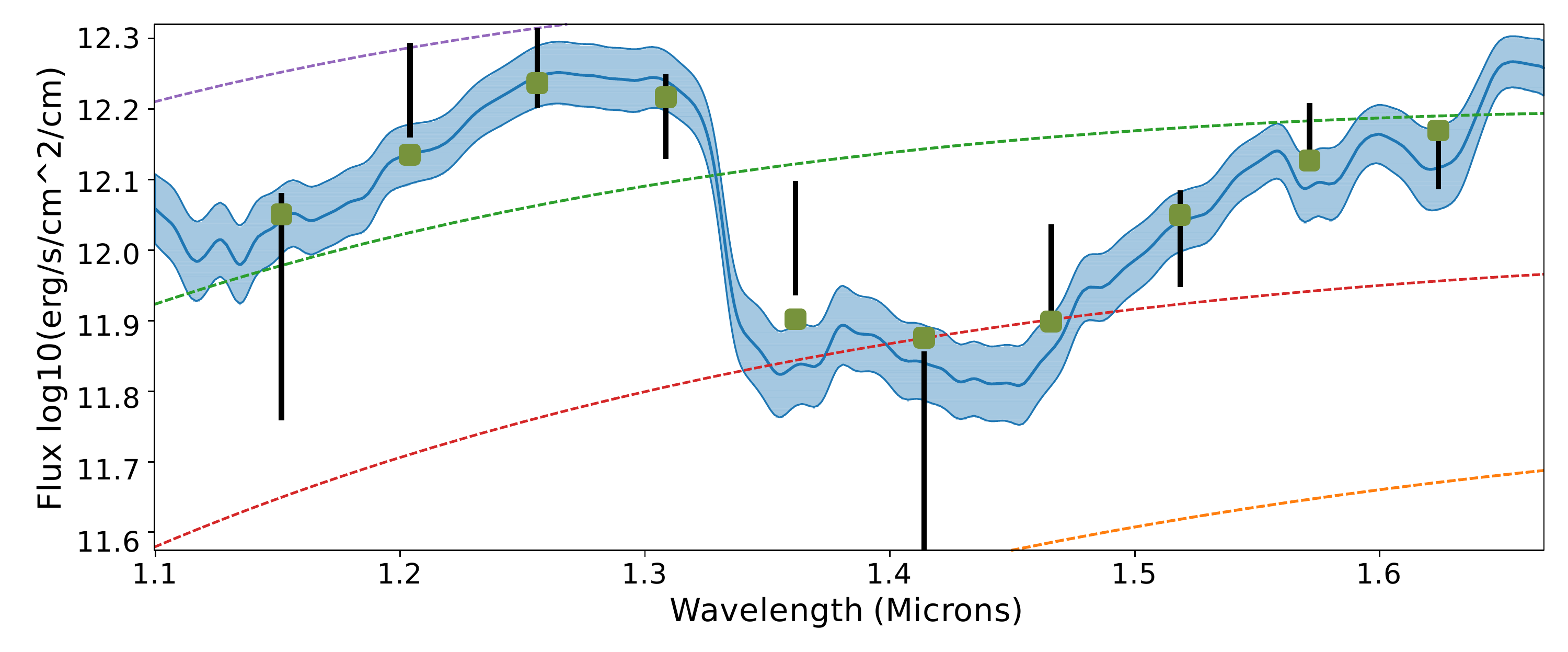}
	\caption{Retrieved planet spectrum with its 1-$\sigma$ uncertainty region, convolved to the HST/WFC3 instrument resolution of R$\sim$130, compared to observations of HD 209458b  from \cite{line:2016}. Green points represent the binned median retrieved spectrum. The bottom figure is a zoomed in version of the top, highlighting the 1.4 micron H$_2$O feature from HST/WFC3 observations. \label{fig:209spec}}
\end{figure}

\begin{figure}
	\center   
	\includegraphics[width=3.5in]{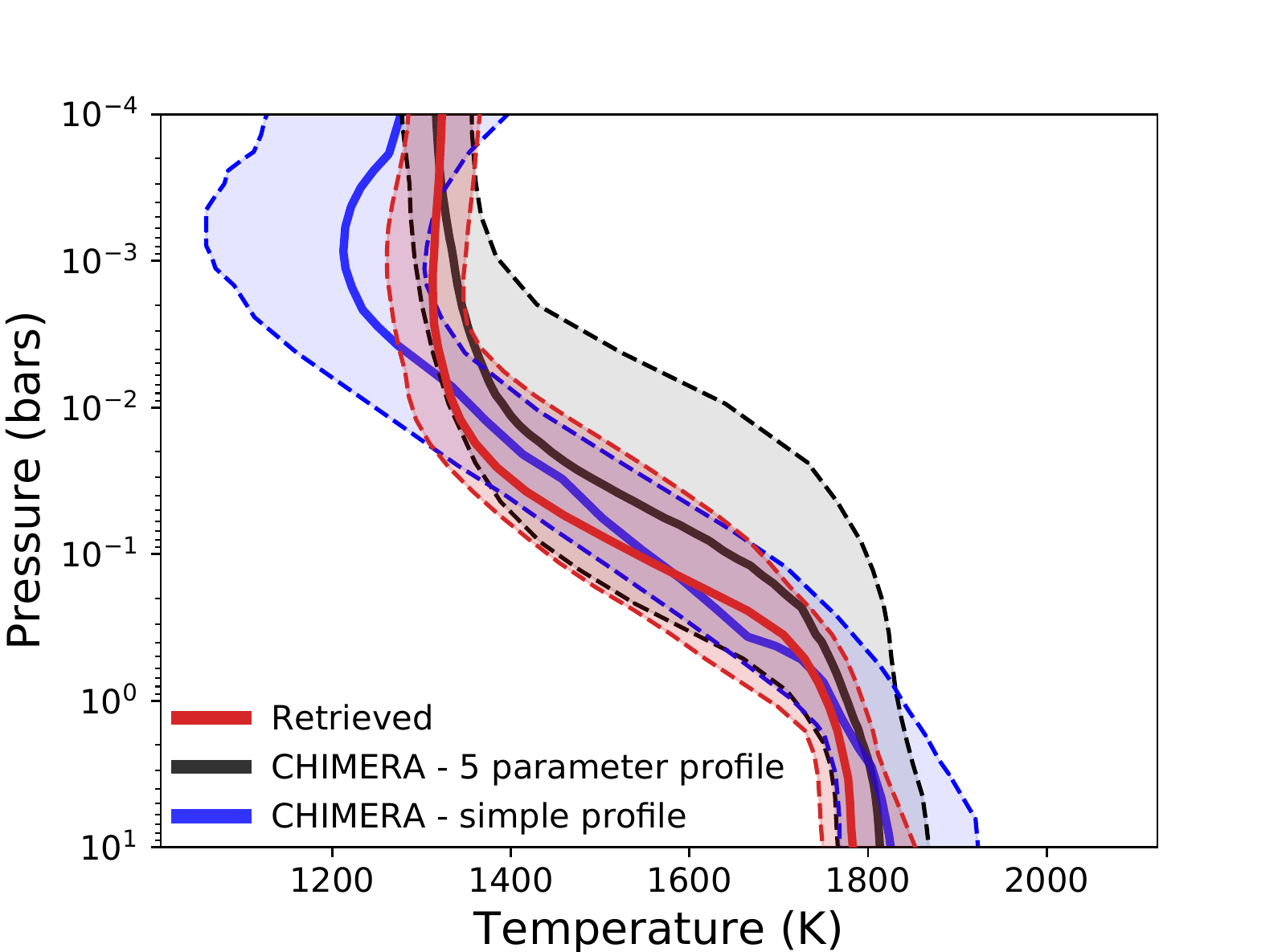}
	\caption{Retrieved temperature profile of the HD 209458b data compared with the median retrieved temperature profile from \citep{line:2016}. Our median retrieved temperature profile agrees with those from the CHIMERA retrieval suite, though our 1-$\sigma$ range is much tighter. This may be because of differences in line lists and statistical framework. \label{fig:209TP}}
\end{figure}

\begin{figure}
	\center  
	\includegraphics[width=3.0in]{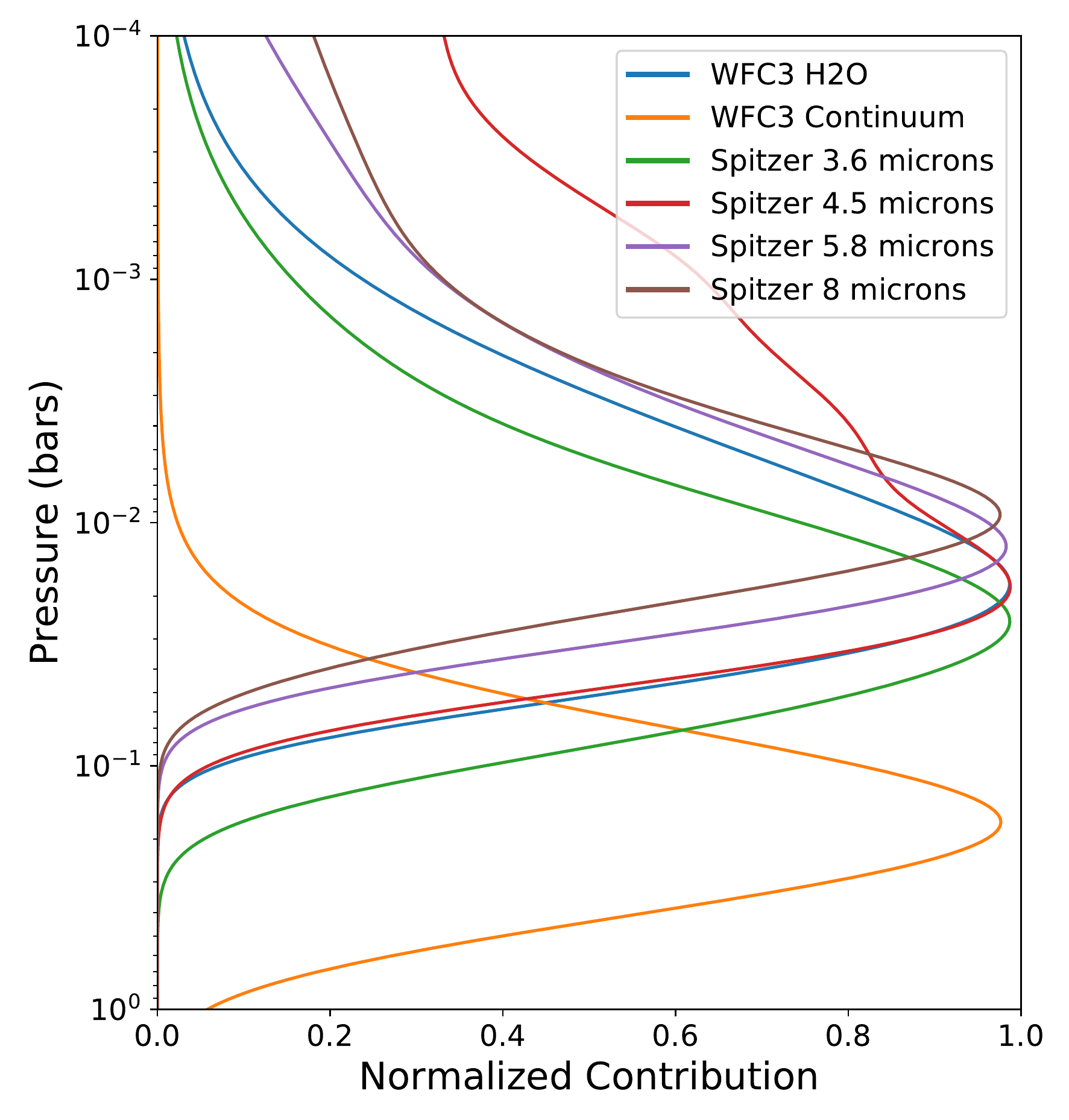}
	\caption{Contribution functions for HD 209458b's dayside atmosphere at 1.25, 1.4, 3.6, and 4.5 \microns{}, corresponding to outside the 1.4 \microns{} H$_2$O feature, inside the 1.4 \microns{} H$_2$O feature, and channel 1, 2, 3, and 4 from \spitzer{}, respectively. \label{fig:hd209_cf}}
\end{figure}

\begin{figure*}
	\center   
	\includegraphics[width=6in]{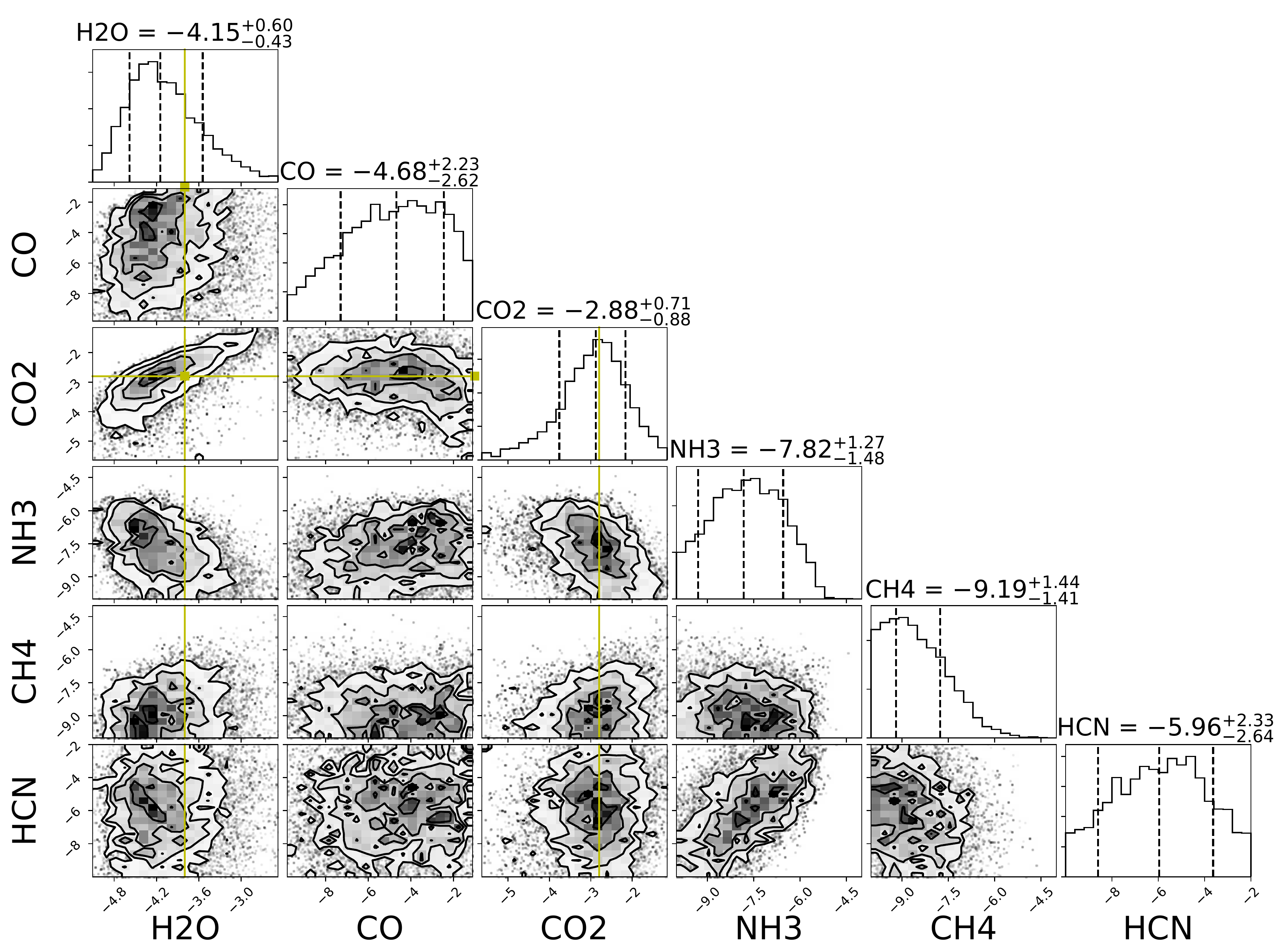}
	\caption{Posterior distributions for the molecular abundances from PETRA's retrieval of HD 209458b \citep{line:2016}. Gold lines represent the approximate maximum of the posterior from \citep{line:2016} for the two species with obvious maxima, H$_2$O and CO$_2$.\label{fig:209mol}}
\end{figure*}

In order to further compare PETRA with other retrieval suites, we retrieved the temperature profile and molecular abundances of hot Jupiter HD 209458b using HST/WFC3 and \textit{Spitzer} data from \cite{line:2016}. This same data was analyzed in \cite{line:2016} with the same CHIMERA retrieval suite as WASP-43b above and was found to show evidence for a non-inverted atmosphere with an approximately solar abundance of H$_2$O. For this retrieval, we use the BT2 line list with Voigt line profiles, which should most closely match the CHIMERA retrieval.

Our retrieval results agree quite well with the CHIMERA retrieval. Figure~\ref{fig:209spec} shows the retrieved spectrum compared to the observations from \cite{line:2016}. The median retrieved spectrum has \chisq{} = 15.2 with 14 data points, leading to a \chisq{} per data point of 1.08. As with WASP-43b, PETRA fits the data as well as CHIMERA, which obtains a \chisq{} per data point of 1.03.

Figure~\ref{fig:209TP} shows that PETRA's retrieved median temperature profile agrees quite well with CHIMERA's, but is more tightly constrained than the CHIMERA profile using the same parameterization \deleted{as the PETRA retrieval }(that of \cite{parmentier:2013}). The PETRA profile is most similar to the CHIMERA retrieval using the simplified PT profile, which consists of a deep isothermal region and 2 ``T linear-in-P" regions. This simplified CHIMERA profile also exhibits a more constrained profile compared to the CHIMERA retrieval with the 5-parameter \cite{parmentier:2013} parameterization.  The CHIMERA retrieval also utilized nested sampling rather than DEMC, which may explain some of the differences with the PETRA retrieval. 

Contribution function profiles are plotted in Figure~\ref{fig:hd209_cf}, demonstrating that pressures between 5 and 100 mbar are constrained by the observations. As with WASP-43b, all constraints on the temperature profile outside this range are a consequence of the parameterization of the temperature profile.

Figure~\ref{fig:209mol} shows that the molecular abundance constraints PETRA retrieves are also well within 2-$\sigma$ of CHIMERA's retrieved constraints, despite the different statistical framework and  different line lists being used between the two suites. \cite{line:2016} uses HITEMP H$_2$O, CO, and CO$_2$ \citep{rothman:2010}, while we use CO$_2$ from \citep{rothman:2008}, CO from \cite{goorvitch:1994}, and H$_2$O from \cite{barber:2006}. 

\added{The inclusion of H$_2$O is strongly favored with change in the BIC of 35. When we neglected CO and CO$_2$, PETRA attempted to compensating by increasing the HCN abundance to a log$_{10}$(VMR) = 2.7. Because HCN also absorbs 1.53 \microns{}, this also resulted in the H$_2$O abundance decreasing by about an order of magnitude. When CO, CO$_2$, and HCN are all ignored, we calculate a BIC of 5.06, indicating positive evidence for absorption by a carbon species in the atmosphere of HD209458b, in agreement with \cite{line:2016}.}

\section{Characterizing Ultra-hot Jupiters with Retrievals of H$^{-}$}\label{sec:hminus}

Ultra-hot Jupiters are among the most ideal targets to observe because of their hot, bright daysides, their inflated radii, and their short periods. However, ultra-hot Jupiter atmospheres are often hot enough to dissociate the very molecules we hope to observe. The absence of molecules is exacerbated by the fact that H$^{-}$ \replaced{continuous}{bound-free and free-free} opacity becomes significant at temperatures above 2500 K, which will move the photosphere to lower pressures and mask the spectral features of molecules like H$_2$O \citep{arcangeli:2018,parmentier:2018,lothringer:2018b,kitzmann:2018}. This phenomenon is thought to explain the absence of H$_2$O in WASP-12b, WASP-18b, WASP-103, and HAT-P-7b \citep{arcangeli:2018,kreidberg:2018,mansfield:2018}.

In order to characterize the atmospheres of ultra-hot Jupiters, we must look to non-molecular spectral signatures. While H$^{-}$ opacity mutes molecular spectral features, it also has the potential to help characterize ultra-hot Jupiter atmospheres. H$^{-}$ opacity is dominant at the temperatures and pressures in ultra-hot Jupiters and, importantly, is non-grey, which means that we can use H$^{-}$ to probe different pressures in ultra-hot Jupiters. In fact, H$^{-}$ opacity increases steadily with increasing wavelength from its minimum at about 1.6 \microns{} because of its free-free interactions:

\begin{equation}
h\nu + \rm{H}^- \leftrightarrow \rm{H} + e^-
\end{equation}

\noindent\citep{wildt:1939}. Similarly, H$^{-}$ opacity increases towards short wavelengths from the minimum at 1.6 \microns{} until 0.85 \microns{} because of its bound-free interaction

\begin{equation}\label{eq:ff}
h\nu + \rm{H} + e^- \leftrightarrow \rm{H} + e^-
\end{equation}

\noindent \citep{pannekoek:1931}. Fortunately, this near-IR wavelength range will be explored to great precision with JWST.

\added{Taken a step further, we can use H$^-$ to constrain the $e^-$ density. H$^{-}$ opacity will scale with the product of the $e^{-}$ and H densities in LTE:}

\begin{equation}\label{eq:hminus_opac2}
\alpha_{H^{-}} = n_{H} n_{e^{-}} \sigma_{H^{-}}
\end{equation}
	
\added{	\noindent where $\alpha_{H^{-}}$ is the absorption coefficient for H$^{-}$, $n_{H}$ is the number density of atomic H, $n_{e^{-}}$ is the $e^-$ number density, and $\sigma_{H^{-}}$ is the cross-section for H$^{-}$ in cm$^5$ from \cite{john:1988}. 
	$n_{H}$ will remain relatively constant with height at pressures above 1~$\mu$bar, so we chose to use its equilibrium abundance rather than allow it to be a free parameter. Note that neither Equation~\ref{eq:ff} nor \ref{eq:hminus_opac2} contain the actual abundance of H$^{-}$, as it is ill-defined in the free-free interaction}

\subsection{KELT-9b}

KELT-9b is the hottest known Jovian planet with a dayside-redistribution equilibrium temperature of about 4500 K \citep{gaudi:2017}, \replaced{about}{over} 1000~K hotter than the next hottest Jovian planet, WASP-33b\footnote{https://exo.mast.stsci.edu/}. KELT-9b provides an ideal example to explore ultra-hot Jupiter characterization through H$^{-}$. In a planet like KELT-9b, the brightness temperature observed in secondary eclipse can vary by nearly 1000~K between 2 and 10 \microns{} due to H$^{-}$ opacity \citep[see Figure~\ref{fig:k9obs2} and see Figure~15 in][]{lothringer:2018b}. This corresponds to probing about an order of magnitude in pressure between 10 and 100~mbar, useful for determining the presence and magnitude of temperature inversions which are predicted to be ubiquitous in the hottest Jovian planets \citep{lothringer:2018b}. \added{Indeed, a temperature inversion has recently been detected in KELT-9b from ground-based high-resolution observations of neutral Fe \citep{pino:2020}.}

\subsubsection{Retrieval with PETRA}
\begin{figure*}
	\center    \includegraphics[width=6in]{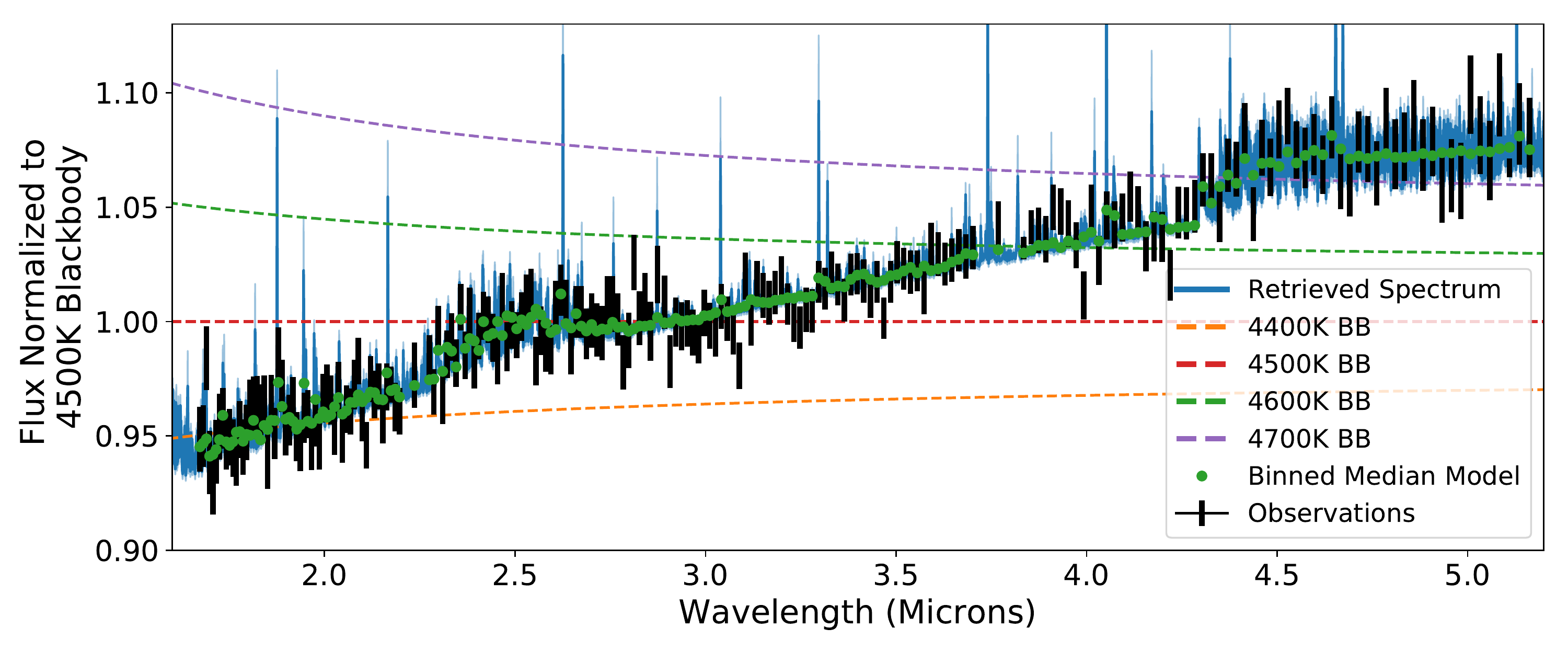}
	\caption{Retrieved median spectrum (blue)\added{ for the retrieval where CO is assumed to be in chemical abundance}, normalized by a 4500 K blackbody and compared to the simulated observations for the KELT-9b retrieval (black). The 1-$\sigma$ uncertainty region is shaded, but small compared to the individual error bars. The green dots represent the binned flux of the median retrieved model. \label{fig:k9obs2}}
\end{figure*}

In order to explore the possibility of retrieving atmospheric properties from ultra-hot Jupiters using H$^{-}$ opacity, we simulate a single secondary eclipse of KELT-9b with both JWST/NIRSPEC/G235H and G395H with PandExo \citep{batalha:2017}. We input to PandExo a fully self-consistent model of KELT-9b, assuming dayside-heat redistribution, from \citep{lothringer:2018b}. The use of a self-consistent model helps to test the retrieval forward model\added{ parameterizations and} assumptions described below.

\begin{figure}
	\center  
	\includegraphics[width=3.0in]{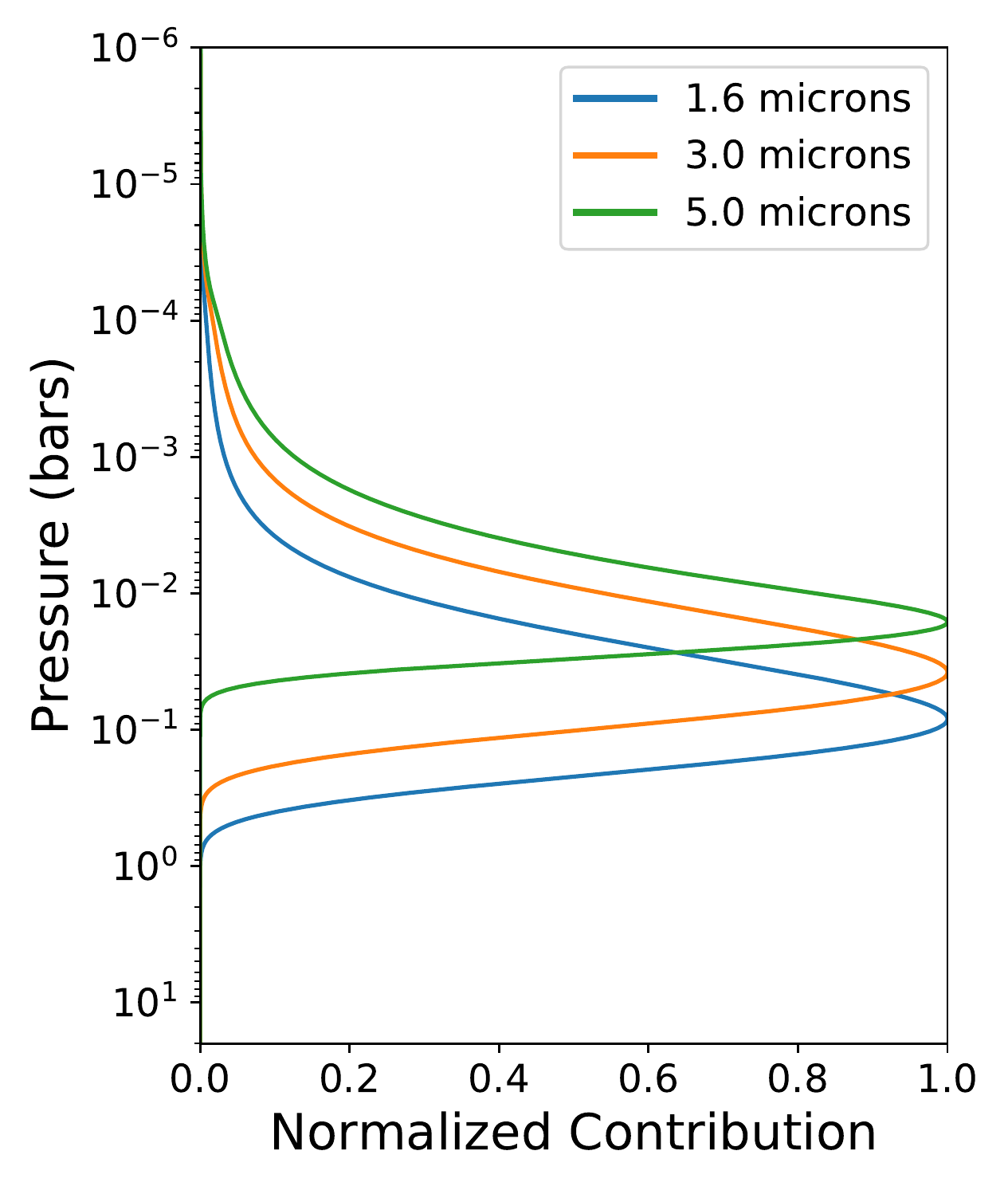}
	\caption{Contribution functions for KELT-9's dayside atmosphere at 1.6, 3.0, and 5.0 \microns{} from the median retrieved model. \label{fig:k9_cf}}
\end{figure}

The G235H and G395H grisms span 1.67-5.14 \microns{} at resolutions of about R$\sim$2,700 with some overlap between 2.87 and 3.05 \microns{}. Shorter wavelength and lower resolution observations are made difficult because of saturation from KELT-9b's bright A0 host star (V=7.6). The broad wavelength coverage afforded by JWST is essential to maximizing brightness temperature contrasts since H$^{-}$ opacity increases steadily but slowly across the wavelength range. Figure~\ref{fig:k9obs2} shows the simulated observations with blackbodies over-plotted for reference, demonstrating that the brightness temperature of KELT-9b changes by over 300 K in the G235H and G395H wavelength region. The contribution functions plotted in Figure~\ref{fig:k9_cf} show that about an order of magnitude in pressure between 10 and 100 mbar is probed between 1.6 and 5 \microns{}.

\begin{figure}
	\center    \includegraphics[width=3.5in]{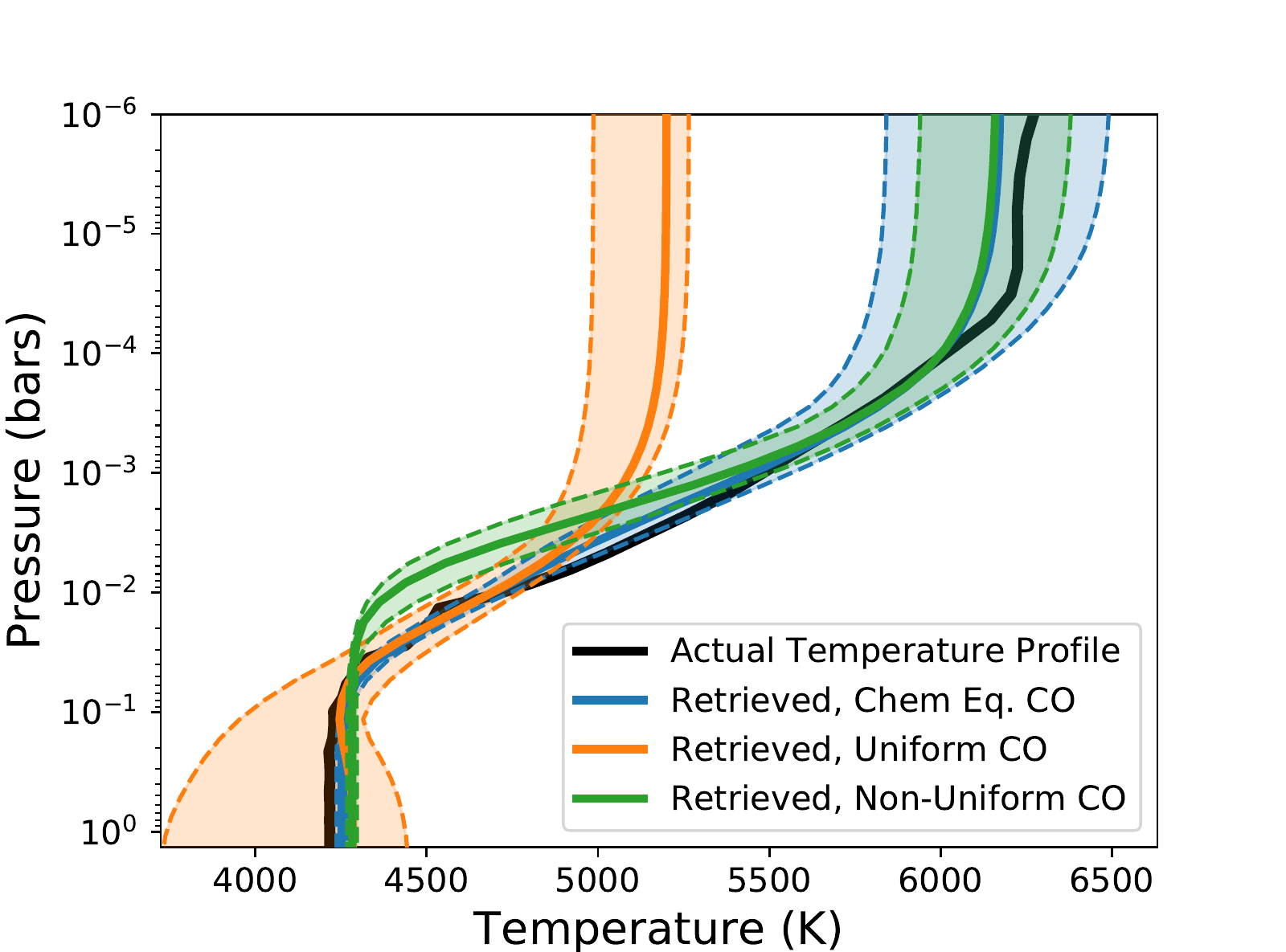}
	\caption{\replaced{Retrieved constraints on the temperature structure of KELT-9b from simulated JWST/NIRSpec data compared to the input temperature profile. The shaded region represent 1-$\sigma$ uncertainties.}{Simulated constraints on the temperature structure of KELT-9b for the three retrieval scenarios, where CO is assumed to be in chemical equilibrium, have a vertically-uniform abundance, or a parameterized evidence of a vertically non-uniform H2O abundance  abundance.} \label{fig:k9tp}}
\end{figure}

\begin{figure*}
	\center    \includegraphics[width=6in]{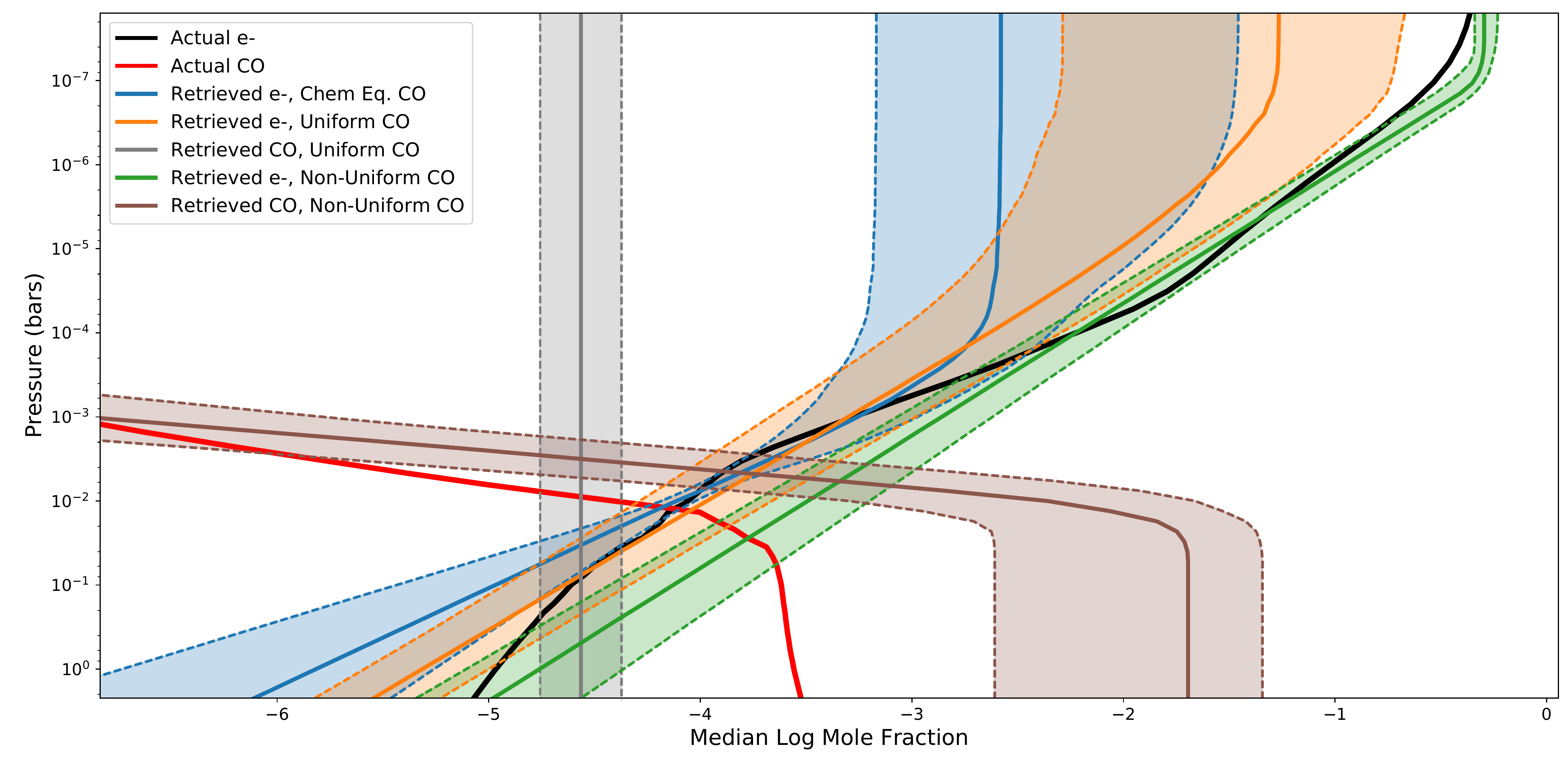}
	\caption{\replaced{Retrieved constraints on the $e^-$ density of KELT-9b from simulated JWST/NIRSpec data compared to the input . The shaded region represent 1-$\sigma$ uncertainties.}{Simulated constraints on the abundances of $e^-$ and CO for the three KELT-9b retrieval scenarios, where CO is assumed to be in chemical equilibrium, have a vertically-uniform abundance, or a parameterized evidence of a vertically non-uniform H2O abundance abundance.} \label{fig:k9e}}
\end{figure*}

We then used PETRA to retrieve atmospheric properties from the simulated KELT-9b observations. We chose to retrieve \deleted{both} the temperature structure\added{, the CO abundance,} and the e$^{-}$ density\added{ using H$^-$ opacity as a proxy, while holding other abundances to their chemical equilibrium value}. \added{Besides H$^-$, our models suggest CO is the only opacity source detectable at low-resolution.} \deleted{As above, we use the temperature profile parameterization from \cite{parmentier:2014} and \cite{line:2013}. To retrieve e$^{-}$, we use the H$^-$ opacity as a proxy, since H$^{-}$ opacity will scale with the product of the $e^{-}$ and H densities in LTE:
\begin{equation}\label{eq:hminus_opac}
\alpha_{H^{-}} = n_{H} n_{e^{-}} \sigma_{H^{-}}
\end{equation}}

\deleted{\noindent where $\alpha_{H^{-}}$ is the absorption coefficient for H$^{-}$, $n_{H}$ is the number density of atomic H, $n_{e^{-}}$ is the $e^-$ number density, and $\sigma_{H^{-}}$ is the cross-section for H$^{-}$ in cm$^5$ from \cite{john:1988}. 
$n_{H}$ will remain relatively constant with height at pressures above 1~$\mu$bar, so we chose to use its equilibrium abundance rather than use it as a free parameter. H$^{-}$ opacity is predicted to be by far the dominant opacity source at IR wavelengths for KELT-9b, so we chose to hold other abundances to their chemical equilibrium values.} 

\replaced{The $e^{-}$ density, $n_{e^{-}}$, will likely not be uniform with height in KELT-9b's atmosphere. It is therefore necessary to parameterize $n_{e^{-}}$ as a function of pressure, as described in Section~\ref{sec:nonuni}. PETRA therefore uses five free parameters to describe the temperature structure and three free parameters to describe $n_{e^{-}}$ for a total of eight free parameters.}{Molecular, atomic, and ion abundances will likely not be uniform with altitude in KELT-9b's atmosphere due to thermal and photo-ionization, as well as the thermal dissociation of molecules. It is therefore necessary to parameterize retrieved chemical abundances. We ran three different scenarios where \textbf{1)} CO is assumed to be in chemical equilibrium, \textbf{2)} the CO abundance is retrieved assuming a vertically-uniform abundance, and \textbf{3)} the CO abundance is retrieved using the evidence of a vertically non-uniform H2O abundance parameterization described in Section~\ref{sec:nonuni}. For each of these scenarios, the $e^-$ abundance is also retrieved, assuming a non-uniform abundance}.

Figure~\ref{fig:k9obs2} shows the retrieved median spectrum \added{for the retrieval where CO is assumed to be in chemical equilibrium }with the 1-$\sigma$ uncertainty region compared to the simulated observations. The uncertainties on the flux for a given wavelength point are much smaller than an individual bin's observational uncertainty.

Figure~\ref{fig:k9tp} and \ref{fig:k9e} shows the \replaced{resulting fit to}{retrieved constraints on} the temperature structure and \replaced{$n_{e^{-}}$}{chemical abundances} (in the form of volume mixing ratio)\replaced{ from the retrieval, respectively}{, respectively, for each retrieval}. The temperature structure is tightly constrained by the retrieval, clearly identifying the strong temperature inversion\added{, however accurate temperatures are only retrieved near the photosphere. Further, the retrieval where CO is parameterized as having a non-uniform abundance (i.e., the model with the most free parameters), actually has the least accurate fit and is biased towards a lower-pressure photosphere}. \deleted{Additionally, the retrieval is able to constrain both the value of $n_{e^{-}}$ at the photosphere, as well as its slope with pressure, to be in agreement with the self-consistent input model. The fact that $n_{e^{-}}$ is constrained above and below the photosphere is simply a consequence of the parameterization, so one should consider only $n_{e^{-}}$ near the photosphere to be physically measured. Also note that the input model for the simulated data did not use a parameterization for the temperature structure or $n_{e^{-}}$, indicating that these parameterizations are appropriate enough to fit observations. Nonetheless, more physically motivated parameterizations, especially for $n_{e^{-}}$, could be worth exploring.}

\added{Figure~\ref{fig:k9e} shows the retrieved chemical abundances compared to the actual abundances of $e^-$ and CO in the input model. When CO is assumed to be in chemical equilibrium, accurate constraints on the $e^-$ density can be obtained. Similarly, if CO is assumed to have a vertically-uniform abundance, accurate $e^-$ and CO abundances are retrieved, albeit with the foreknowledge that the uniform CO abundance can only be an approximation. Interestingly, however, when both CO and $e^-$ are retrieved using the non-uniform parameterization, the abundances are biased to higher values, which help place the photosphere at a lower pressure as described above.}

\begin{figure*}
	\center    \includegraphics[width=6.5in]{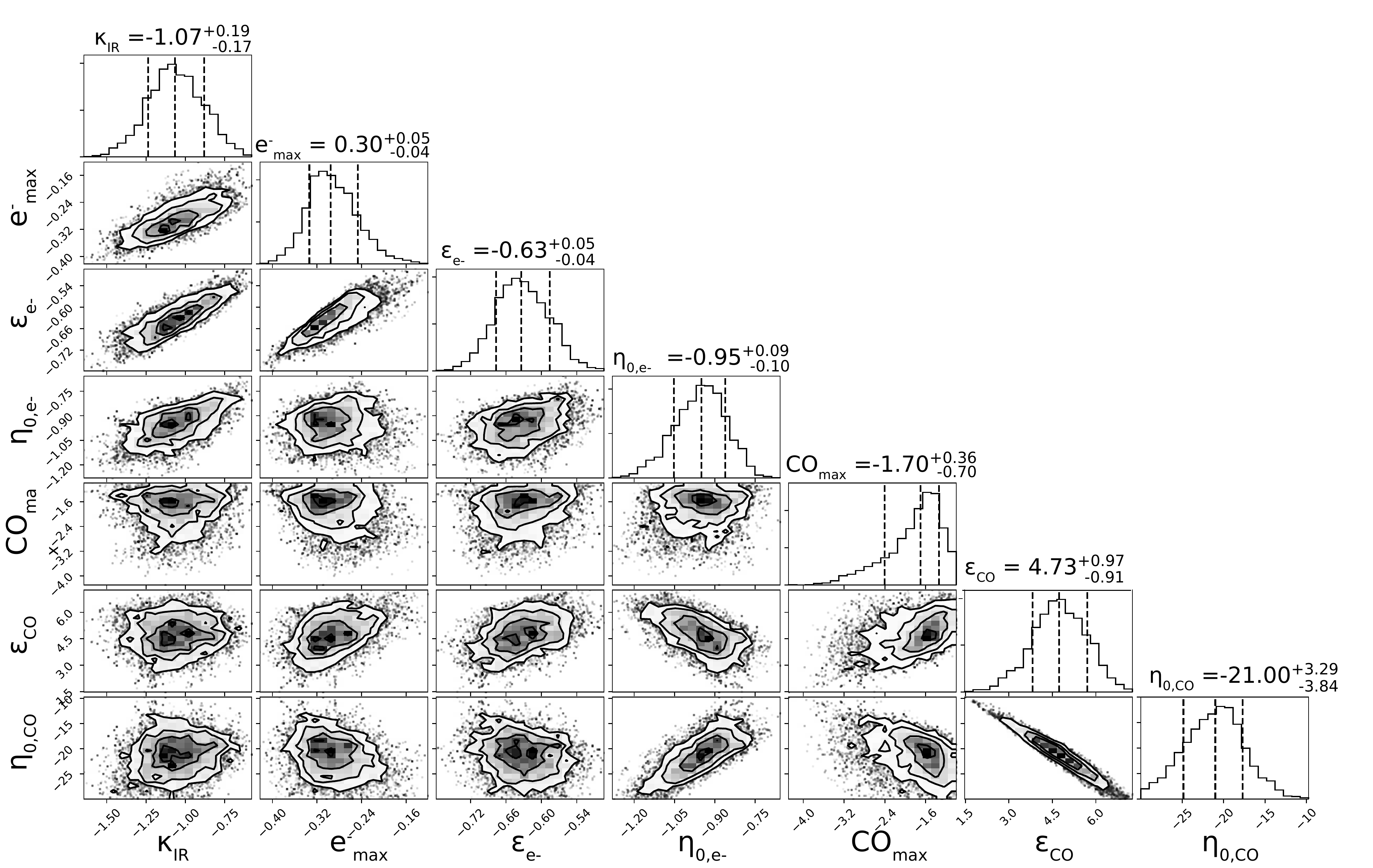}
	\caption{\added{Posterior distributions for the KELT-9b retrieval where a evidence of a vertically non-uniform H2O abundance CO abundance parameterization was used. For clarity, most temperature structure parameters were not included. Note the correlations between $\kappa_{IR}$, e$^{-}_{max}$, and $\eta_{0,e^-}$, as well as the correlations between $\eta_{0,e^-}$, $\eta_{0,CO}$, and $\epsilon_{0,CO}$. \label{fig:k9nonunicorner}}}
\end{figure*}

\added{The reason for the behavior seen when CO and $e^-$ are retrieved non-uniformly is due to the fact that there is a degeneracy between the photospheric level and the chemical abundances. In the \cite{parmentier:2014} temperature structure parameterization, the pressure-level of the photosphere is controlled through the $\kappa_{IR}$ parameter which physically represents the mean opacity of the IR and sets the relation between the optical depth and pressure. Figure~\ref{fig:k9nonunicorner} shows the posterior distribution for the chemical abundance parameters and the $\kappa_{IR}$ parameter. A clear correlation exists between $\kappa_{IR}$ and the $e^-$ abundance parameters. There is also some degeneracy between the slope of the $e^-$ density, $\eta_{0,e^-}$, and the CO abundance parameters.}

\added{A degeneracy between the abundance parameters and $\kappa_{IR}$ exits because there is no non-retrieved opacity that can set the photospheric level. In the case where CO is assumed to be in chemical equilibrium, the relation between the brightness temperature at 2.3 and 4.5 microns and the CO opacity at those wavelengths effectively anchors the temperature structure. Indeed, the degeneracy is broken in the retrieval where CO is in chemical equilibrium and can be used as such an anchor.}

\added{This situation is similar to the well-known degeneracy in transit spectra between the reference radius and pressure \citep{griffith:2014,heng:2017}. This degeneracy can be broken through information about the scale height from the Rayleigh scattering slope \citep{benneke:2012,line:2016b} or from the continuum level from H$_2$ CIA opacity \citep{welbanks:2019}. H$^-$ opacity would serve to similarly break the degeneracy in for ultra-hot Jupiters. In our retrieved emission spectrum, however, we retrieved both the continuum opacity and the CO abundance and had nothing to anchor the photospheric level of the atmosphere. An assumption, like chemical equilibrium of CO, provides this anchor and breaks the correlation.}

The power of H$^{-}$ retrievals lies not only in their ability to constrain ultra-hot Jupiter temperature structures, but also in the retrieval's measurement of $n_{e^{-}}$. A deviation in $n_{e^{-}}$ from chemical equilibrium could identify and measure the effects of photoionization in these highly irradiated atmospheres. Additionally, a direct measurement of the ion density will provide a path forward to understanding magneto-hydrodynamic (MHD) effects in ultra-hot Jupiters, which are likely affecting atmospheric circulation through magnetic drag forces \citep{rogersandtad:2014,tad:2016,rogers:2017b}. The combination of ion density measurements through H$^{-}$ retrievals along with wind speed measurements from high-dispersion spectroscopy \citep[e.g.,]{brogi:2016} and day-night temperature contrasts from phase curves will provide direct insight into MHD effects. 

\added{We also note that rather than using H$^{-}$ as a free variable, it can be used in retrievals of metallicity that assume chemical equilibrium \citep{arcangeli:2018}. When used in this fashion, H$^{-}$ can help constrain the metallicity because it will depend on the number of free electrons, which will be supplied from metals like Na, K, and Ca. This would not account for the effect of photoionization, however.}

\section{Conclusion}

We have introduced the PHOENIX ExoplaneT Retrieval Algorithm, or PETRA, which is a retrieval framework built around the PHOENIX atmosphere model. PETRA is flexible enough to retrieve atmospheric properties from observations of exoplanets, brown dwarfs, and even stars for a variety of situations, including transit and eclipse spectroscopy, as well as self-luminous/directly-imaged objects.

We have validated PETRA by retrieving properties from data simulated from PHOENIX models, for which the correct atmospheric properties are known. PETRA is successfully able to retrieve these correct properties. We have further validated PETRA against previous retrieval results for actual HST/WFC3 and \textit{Spitzer} observations of WASP-43b and HD 209458b, showing that PETRA can agree with well-tested retrievals like CHIMERA and ATMO. For WASP-43b, we also investigated the effect that different H$_2$O line lists and line shape treatments have on retrieved quantities, showing significant differences especially in the retrieved temperature profiles. \added{Our results are similar to those of \citep{barstow:2020}, which demonstrated comparable agreement between retrieval suites and the importance of line broadening treatment.}

We then used PETRA to demonstrate a novel technique to characterize ultra-hot Jupiters through retrieving H$^{-}$. Retrievals using current techniques (i.e., retrieving molecular abundances) are hampered by the thermal dissociation of molecules and presence of H$^{-}$ opacity at high temperatures. Using simulated JWST/NIRSPEC data, we showed that retrieving the H$^{-}$ can constrain the temperature structure and $e^{-}$ density, providing a unique path forward towards characterizing the hottest Jovian exoplanets. \added{We discovered that an assumption about the CO opacity is critical to using H$^{-}$ in a retrieval due to correlations between the parameters that set the photospheric level of the temperature structure and the chemical abundances.} \added{We note that ground-based high-resolution studies will be insensitive to the continuous opacity of H$^{-}$, meaning that JWST may be the only facility capable of this experiment until the \textit{ARIEL} mission \citep{pascale:2020}.}

PETRA will be continually improved. We plan to incorporate PHOENIX's cloud modeling capabilities into PETRA to perform retrievals of cool objects. We will also improve the statistical framework by implementing nested sampling in order to robustly perform model complexity comparison. In the future, we will apply PETRA to a variety of situations, from transmission spectra of sub-Neptunes to high resolution spectroscopy of giant planets.

\acknowledgments
We thank the anonymous referee for helpful comments that improved the manuscript. We also thank Ian Crossfield, Tommi Koskinen, Roger Yelle, and Andrew Youdin for constructive feedback and discussion. This research has made use of the NASA Astrophysics Data System and the NASA Exoplanet Archive, which is operated by the California Institute of Technology, under contract with the National Aeronautics and Space Administration under the Exoplanet Exploration Program. Allocation of computer time from the UA Research Computing High Performance Computing (HPC) at the University of Arizona is also gratefully acknowledged. 
       
\software{Corner \citep{corner}, PandExo \citep{batalha:2017}, Astropy \citep{astropy:2013}}
    
\vspace{15pt}
\bibliographystyle{aasjournal}

\begin{thebibliography}{}
	\expandafter\ifx\csname natexlab\endcsname\relax\def\natexlab#1{#1}\fi
	\providecommand{\url}[1]{\href{#1}{#1}}
	\providecommand{\dodoi}[1]{doi:~\href{http://doi.org/#1}{\nolinkurl{#1}}}
	\providecommand{\doeprint}[1]{\href{http://ascl.net/#1}{\nolinkurl{http://ascl.net/#1}}}
	\providecommand{\doarXiv}[1]{\href{https://arxiv.org/abs/#1}{\nolinkurl{https://arxiv.org/abs/#1}}}
	
	\bibitem[{{Allard} {et~al.}(2011){Allard}, {Homeier}, \&
		{Freytag}}]{allard:2010}
	{Allard}, F., {Homeier}, D., \& {Freytag}, B. 2011, in Astronomical Society of
	the Pacific Conference Series, Vol. 448, 16th Cambridge Workshop on Cool
	Stars, Stellar Systems, and the Sun, ed. C.~{Johns-Krull}, M.~K. {Browning},
	\& A.~A. {West}, 91.
	\newblock \doarXiv{1011.5405}
	
	\bibitem[{{Arcangeli} {et~al.}(2018){Arcangeli}, {D{\'e}sert}, {Line}, {Bean},
		{Parmentier}, {Stevenson}, {Kreidberg}, {Fortney}, {Mansfield}, \&
		{Showman}}]{arcangeli:2018}
	{Arcangeli}, J., {D{\'e}sert}, J.-M., {Line}, M.~R., {et~al.} 2018, \apjl, 855,
	L30, \dodoi{10.3847/2041-8213/aab272}
	
	\bibitem[{{Astropy Collaboration} {et~al.}(2013){Astropy Collaboration},
		{Robitaille}, {Tollerud}, {Greenfield}, {Droettboom}, {Bray}, {Aldcroft},
		{Davis}, {Ginsburg}, {Price-Whelan}, {Kerzendorf}, {Conley}, {Crighton},
		{Barbary}, {Muna}, {Ferguson}, {Grollier}, {Parikh}, {Nair}, {Unther},
		{Deil}, {Woillez}, {Conseil}, {Kramer}, {Turner}, {Singer}, {Fox}, {Weaver},
		{Zabalza}, {Edwards}, {Azalee Bostroem}, {Burke}, {Casey}, {Crawford},
		{Dencheva}, {Ely}, {Jenness}, {Labrie}, {Lim}, {Pierfederici}, {Pontzen},
		{Ptak}, {Refsdal}, {Servillat}, \& {Streicher}}]{astropy:2013}
	{Astropy Collaboration}, {Robitaille}, T.~P., {Tollerud}, E.~J., {et~al.} 2013,
	\aap, 558, A33, \dodoi{10.1051/0004-6361/201322068}
	
	\bibitem[{{Barber} {et~al.}(2006){Barber}, {Tennyson}, {Harris}, \&
		{Tolchenov}}]{barber:2006}
	{Barber}, R.~J., {Tennyson}, J., {Harris}, G.~J., \& {Tolchenov}, R.~N. 2006,
	\mnras, 368, 1087, \dodoi{10.1111/j.1365-2966.2006.10184.x}
	
	\bibitem[{{Barman} {et~al.}(2001){Barman}, {Hauschildt}, \&
		{Allard}}]{barman:2001}
	{Barman}, T.~S., {Hauschildt}, P.~H., \& {Allard}, F. 2001, \apj, 556, 885,
	\dodoi{10.1086/321610}
	
	\bibitem[{{Barman} {et~al.}(2011){Barman}, {Macintosh}, {Konopacky}, \&
		{Marois}}]{barman:2011}
	{Barman}, T.~S., {Macintosh}, B., {Konopacky}, Q.~M., \& {Marois}, C. 2011,
	\apj, 733, 65, \dodoi{10.1088/0004-637X/733/1/65}
	
	\bibitem[{{Barstow} {et~al.}(2017){Barstow}, {Aigrain}, {Irwin}, \&
		{Sing}}]{barstow:2017}
	{Barstow}, J.~K., {Aigrain}, S., {Irwin}, P.~G.~J., \& {Sing}, D.~K. 2017,
	\apj, 834, 50, \dodoi{10.3847/1538-4357/834/1/50}
	
	\bibitem[{{Barstow} {et~al.}(2020){Barstow}, {Changeat}, {Garland}, {Line},
		{Rocchetto}, \& {Waldmann}}]{barstow:2020}
	{Barstow}, J.~K., {Changeat}, Q., {Garland}, R., {et~al.} 2020, \mnras, 493,
	4884, \dodoi{10.1093/mnras/staa548}
	
	\bibitem[{{Barstow} \& {Heng}(2020)}]{barstow:2020b}
	{Barstow}, J.~K., \& {Heng}, K. 2020, arXiv e-prints, arXiv:2003.14311.
	\newblock \doarXiv{2003.14311}
	
	\bibitem[{{Batalha} {et~al.}(2017){Batalha}, {Mandell}, {Pontoppidan},
		{Stevenson}, {Lewis}, {Kalirai}, {Earl}, {Greene}, {Albert}, \&
		{Nielsen}}]{batalha:2017}
	{Batalha}, N.~E., {Mandell}, A., {Pontoppidan}, K., {et~al.} 2017, \pasp, 129,
	064501, \dodoi{10.1088/1538-3873/aa65b0}
	
	\bibitem[{{Benneke} \& {Seager}(2012)}]{benneke:2012}
	{Benneke}, B., \& {Seager}, S. 2012, \apj, 753, 100,
	\dodoi{10.1088/0004-637X/753/2/100}
	
	\bibitem[{{Brogi} {et~al.}(2016){Brogi}, {de Kok}, {Albrecht}, {Snellen},
		{Birkby}, \& {Schwarz}}]{brogi:2016}
	{Brogi}, M., {de Kok}, R.~J., {Albrecht}, S., {et~al.} 2016, \apj, 817, 106,
	\dodoi{10.3847/0004-637X/817/2/106}
	
	\bibitem[{{Brogi} {et~al.}(2017){Brogi}, {Line}, {Bean}, {D{\'e}sert}, \&
		{Schwarz}}]{brogi:2017}
	{Brogi}, M., {Line}, M., {Bean}, J., {D{\'e}sert}, J.-M., \& {Schwarz}, H.
	2017, \apjl, 839, L2, \dodoi{10.3847/2041-8213/aa6933}
	
	\bibitem[{Brogi {et~al.}(2019)Brogi, Line, {Brogi}, \& {Line}}]{brogi:2018}
	Brogi, M., Line, M.~R., {Brogi}, M., \& {Line}, M.~R. 2019, \aj, 157, 114,
	\dodoi{10.3847/1538-3881/aaffd3}
	
	\bibitem[{{Burningham} {et~al.}(2017){Burningham}, {Marley}, {Line}, {Lupu},
		{Visscher}, {Morley}, {Saumon}, \& {Freedman}}]{burningham:2017}
	{Burningham}, B., {Marley}, M.~S., {Line}, M.~R., {et~al.} 2017, \mnras, 470,
	1177, \dodoi{10.1093/mnras/stx1246}
	
	\bibitem[{{Evans} {et~al.}(2017){Evans}, {Sing}, {Kataria}, {Goyal}, {Nikolov},
		{Wakeford}, {Deming}, {Marley}, {Amundsen}, {Ballester}, {Barstow},
		{Ben-Jaffel}, {Bourrier}, {Buchhave}, {Cohen}, {Ehrenreich}, {Garc{\'{\i}}a
			Mu{\~n}oz}, {Henry}, {Knutson}, {Lavvas}, {Etangs}, {Lewis},
		{L{\'o}pez-Morales}, {Mandell}, {Sanz-Forcada}, {Tremblin}, \&
		{Lupu}}]{evans:2017}
	{Evans}, T.~M., {Sing}, D.~K., {Kataria}, T., {et~al.} 2017, \nat, 548, 58,
	\dodoi{10.1038/nature23266}
	
	\bibitem[{{Fisher} {et~al.}(2019){Fisher}, {Hoeijmakers}, {Kitzmann},
		{M{\'a}rquez-Neila}, {Grimm}, {Sznitman}, \& {Heng}}]{fisher:2019}
	{Fisher}, C., {Hoeijmakers}, H.~J., {Kitzmann}, D., {et~al.} 2019, arXiv
	e-prints, arXiv:1910.11627.
	\newblock \doarXiv{1910.11627}
	
	\bibitem[{Foreman-Mackey(2016)}]{corner}
	Foreman-Mackey, D. 2016, The Journal of Open Source Software, 1, 24,
	\dodoi{10.21105/joss.00024}
	
	\bibitem[{{Gandhi} \& {Madhusudhan}(2018)}]{gandhi:2018}
	{Gandhi}, S., \& {Madhusudhan}, N. 2018, \mnras, 474, 271,
	\dodoi{10.1093/mnras/stx2748}
	
	\bibitem[{{Gandhi} {et~al.}(2019){Gandhi}, {Madhusudhan}, {Hawker}, \&
		{Piette}}]{gandhi:2019b}
	{Gandhi}, S., {Madhusudhan}, N., {Hawker}, G., \& {Piette}, A. 2019, arXiv
	e-prints, arXiv:1910.14042.
	\newblock \doarXiv{1910.14042}
	
	\bibitem[{{Gaudi} {et~al.}(2017){Gaudi}, {Stassun}, {Collins}, {Beatty},
		{Zhou}, {Latham}, {Bieryla}, {Eastman}, {Siverd}, {Crepp}, {Gonzales},
		{Stevens}, {Buchhave}, {Pepper}, {Johnson}, {Colon}, {Jensen}, {Rodriguez},
		{Bozza}, {Novati}, {D'Ago}, {Dumont}, {Ellis}, {Gaillard}, {Jang-Condell},
		{Kasper}, {Fukui}, {Gregorio}, {Ito}, {Kielkopf}, {Manner}, {Matt}, {Narita},
		{Oberst}, {Reed}, {Scarpetta}, {Stephens}, {Yeigh}, {Zambelli}, {Fulton},
		{Howard}, {James}, {Penny}, {Bayliss}, {Curtis}, {Depoy}, {Esquerdo},
		{Gould}, {Joner}, {Kuhn}, {Labadie-Bartz}, {Lund}, {Marshall}, {McLeod},
		{Pogge}, {Relles}, {Stockdale}, {Tan}, {Trueblood}, \&
		{Trueblood}}]{gaudi:2017}
	{Gaudi}, B.~S., {Stassun}, K.~G., {Collins}, K.~A., {et~al.} 2017, \nat, 546,
	514, \dodoi{10.1038/nature22392}
	
	\bibitem[{{Gelman} \& {Rubin}(1992)}]{gelman:1992}
	{Gelman}, A., \& {Rubin}, D.~B. 1992, Statistical Science, 7, 457,
	\dodoi{10.1214/ss/1177011136}
	
	\bibitem[{{Gibson} {et~al.}(2020){Gibson}, {Merritt}, {Nugroho}, {Cubillos},
		{de Mooij}, {Mikal-Evans}, {Fossati}, {Lothringer}, {Nikolov}, {Sing},
		{Spake}, {Watson}, \& {Wilson}}]{gibson:2020}
	{Gibson}, N.~P., {Merritt}, S., {Nugroho}, S.~K., {et~al.} 2020, arXiv
	e-prints, arXiv:2001.06430.
	\newblock \doarXiv{2001.06430}
	
	\bibitem[{{Goorvitch}(1994)}]{goorvitch:1994}
	{Goorvitch}, D. 1994, \apjs, 95, 535, \dodoi{10.1086/192110}
	
	\bibitem[{Gottwald \& Bovensmann(2011)}]{gottwald:2011}
	Gottwald, M., \& Bovensmann, H. 2011, SCIAMACHY - Exploring the Changing
	Earth's Atmosphere, 1, \dodoi{10.1007/978-90-481-9896-2}
	
	\bibitem[{{Gravity Collaboration} {et~al.}(2020){Gravity Collaboration},
		{Nowak}, {Lacour}, {Molli{\`e}re}, {Wang}, {Charnay}, {van Dishoeck},
		{Abuter}, {Amorim}, {Berger}, {Beust}, {Bonnefoy}, {Bonnet}, {Brandner},
		{Buron}, {Cantalloube}, {Collin}, {Chapron}, {Cl{\'e}net}, {Coud{\'e} Du
			Foresto}, {de Zeeuw}, {Dembet}, {Dexter}, {Duvert}, {Eckart}, {Eisenhauer},
		{F{\"o}rster Schreiber}, {F{\'e}dou}, {Garcia Lopez}, {Gao}, {Gendron},
		{Genzel}, {Gillessen}, {Hau{\ss}mann}, {Henning}, {Hippler}, {Hubert},
		{Jocou}, {Kervella}, {Lagrange}, {Lapeyr{\`e}re}, {Le Bouquin}, {L{\'e}na},
		{Maire}, {Ott}, {Paumard}, {Paladini}, {Perraut}, {Perrin}, {Pueyo}, {Pfuhl},
		{Rabien}, {Rau}, {Rodr{\'\i}guez-Coira}, {Rousset}, {Scheithauer},
		{Shangguan}, {Straub}, {Straubmeier}, {Sturm}, {Tacconi}, {Vincent},
		{Widmann}, {Wieprecht}, {Wiezorrek}, {Woillez}, {Yazici}, \&
		{Ziegler}}]{gravity:2020}
	{Gravity Collaboration}, {Nowak}, M., {Lacour}, S., {et~al.} 2020, \aap, 633,
	A110, \dodoi{10.1051/0004-6361/201936898}
	
	\bibitem[{{Gray}(1992)}]{gray:1992}
	{Gray}, D.~F. 1992, {The observation and analysis of stellar photospheres.},
	Vol.~20 (Cambridge University Press, Cambridge, UK)
	
	\bibitem[{{Griffith}(2014)}]{griffith:2014}
	{Griffith}, C.~A. 2014, Philosophical Transactions of the Royal Society of
	London Series A, 372, 20130086, \dodoi{10.1098/rsta.2013.0086}
	
	\bibitem[{{Hauschildt} {et~al.}(1999){Hauschildt}, {Allard}, \&
		{Baron}}]{hauschildt:1999}
	{Hauschildt}, P.~H., {Allard}, F., \& {Baron}, E. 1999, \apj, 512, 377,
	\dodoi{10.1086/306745}
	
	\bibitem[{{Hauschildt} {et~al.}(1997){Hauschildt}, {Baron}, \&
		{Allard}}]{hauschildt:1997}
	{Hauschildt}, P.~H., {Baron}, E., \& {Allard}, F. 1997, \apj, 483, 390,
	\dodoi{10.1086/304233}
	
	\bibitem[{{Heng} \& {Kitzmann}(2017)}]{heng:2017}
	{Heng}, K., \& {Kitzmann}, D. 2017, \mnras, 470, 2972,
	\dodoi{10.1093/mnras/stx1453}
	
	\bibitem[{{Himes} {et~al.}(2020){Himes}, {Harrington}, {Cobb}, {Gunes Baydin},
		{Soboczenski}, {O'Beirne}, {Zorzan}, {Wright}, {Scheffer}, {Domagal-Goldman},
		\& {Arney}}]{himes:2020}
	{Himes}, M.~D., {Harrington}, J., {Cobb}, A.~D., {et~al.} 2020, arXiv e-prints,
	arXiv:2003.02430.
	\newblock \doarXiv{2003.02430}
	
	\bibitem[{{Hjerting}(1938)}]{hjerting:1938}
	{Hjerting}, F. 1938, \apj, 88, 508, \dodoi{10.1086/144000}
	
	\bibitem[{{Howe} {et~al.}(2017){Howe}, {Burrows}, \& {Deming}}]{howe:2017}
	{Howe}, A.~R., {Burrows}, A., \& {Deming}, D. 2017, \apj, 835, 96,
	\dodoi{10.3847/1538-4357/835/1/96}
	
	\bibitem[{{Irwin} {et~al.}(2008){Irwin}, {Teanby}, {de Kok}, {Fletcher},
		{Howett}, {Tsang}, {Wilson}, {Calcutt}, {Nixon}, \& {Parrish}}]{irwin:2008}
	{Irwin}, P.~G.~J., {Teanby}, N.~A., {de Kok}, R., {et~al.} 2008, \jqsrt, 109,
	1136, \dodoi{10.1016/j.jqsrt.2007.11.006}
	
	\bibitem[{{John}(1988)}]{john:1988}
	{John}, T.~L. 1988, \aap, 193, 189
	
	\bibitem[{{Kitzmann} {et~al.}(2020){Kitzmann}, {Heng}, {Oreshenko}, {Grimm},
		{Apai}, {Bowler}, {Burgasser}, \& {Marley}}]{kitzmann:2020}
	{Kitzmann}, D., {Heng}, K., {Oreshenko}, M., {et~al.} 2020, \apj, 890, 174,
	\dodoi{10.3847/1538-4357/ab6d71}
	
	\bibitem[{{Kitzmann} {et~al.}(2018){Kitzmann}, {Heng}, {Rimmer}, {Hoeijmakers},
		{Tsai}, {Malik}, {Lendl}, {Deitrick}, \& {Demory}}]{kitzmann:2018}
	{Kitzmann}, D., {Heng}, K., {Rimmer}, P.~B., {et~al.} 2018, \apj, 863, 183,
	\dodoi{10.3847/1538-4357/aace5a}
	
	\bibitem[{{Komacek} \& {Showman}(2016)}]{tad:2016}
	{Komacek}, T.~D., \& {Showman}, A.~P. 2016, \apj, 821, 16,
	\dodoi{10.3847/0004-637X/821/1/16}
	
	\bibitem[{{Kreidberg} {et~al.}(2014){Kreidberg}, {Bean}, {D{\'e}sert}, {Line},
		{Fortney}, {Madhusudhan}, {Stevenson}, {Showman}, {Charbonneau},
		{McCullough}, {Seager}, {Burrows}, {Henry}, {Williamson}, {Kataria}, \&
		{Homeier}}]{kreidberg:2014b}
	{Kreidberg}, L., {Bean}, J.~L., {D{\'e}sert}, J.-M., {et~al.} 2014, \apjl, 793,
	L27, \dodoi{10.1088/2041-8205/793/2/L27}
	
	\bibitem[{{Kreidberg} {et~al.}(2015){Kreidberg}, {Line}, {Bean}, {Stevenson},
		{D{\'e}sert}, {Madhusudhan}, {Fortney}, {Barstow}, {Henry}, {Williamson}, \&
		{Showman}}]{kreidberg:2015}
	{Kreidberg}, L., {Line}, M.~R., {Bean}, J.~L., {et~al.} 2015, \apj, 814, 66,
	\dodoi{10.1088/0004-637X/814/1/66}
	
	\bibitem[{{Kreidberg} {et~al.}(2018){Kreidberg}, {Line}, {Parmentier},
		{Stevenson}, {Louden}, {Bonnefoy}, {Faherty}, {Henry}, {Williamson},
		{Stassun}, {Beatty}, {Bean}, {Fortney}, {Showman}, {D{\'e}sert}, \&
		{Arcangeli}}]{kreidberg:2018}
	{Kreidberg}, L., {Line}, M.~R., {Parmentier}, V., {et~al.} 2018, \aj, 156, 17,
	\dodoi{10.3847/1538-3881/aac3df}
	
	\bibitem[{{Lavie} {et~al.}(2017){Lavie}, {Ehrenreich}, {Bourrier}, {Lecavelier
			des Etangs}, {Vidal-Madjar}, {Delfosse}, {Gracia Berna}, {Heng}, {Thomas},
		{Udry}, \& {Wheatley}}]{lavie:2017}
	{Lavie}, B., {Ehrenreich}, D., {Bourrier}, V., {et~al.} 2017, \aap, 605, L7,
	\dodoi{10.1051/0004-6361/201731340}
	
	\bibitem[{{Lee} {et~al.}(2013){Lee}, {Heng}, \& {Irwin}}]{lee:2013}
	{Lee}, J.-M., {Heng}, K., \& {Irwin}, P.~G.~J. 2013, \apj, 778, 97,
	\dodoi{10.1088/0004-637X/778/2/97}
	
	\bibitem[{{Line} {et~al.}(2014{\natexlab{a}}){Line}, {Fortney}, {Marley}, \&
		{Sorahana}}]{line:2014b}
	{Line}, M.~R., {Fortney}, J.~J., {Marley}, M.~S., \& {Sorahana}, S.
	2014{\natexlab{a}}, \apj, 793, 33, \dodoi{10.1088/0004-637X/793/1/33}
	
	\bibitem[{{Line} {et~al.}(2014{\natexlab{b}}){Line}, {Knutson}, {Wolf}, \&
		{Yung}}]{line:2013b}
	{Line}, M.~R., {Knutson}, H., {Wolf}, A.~S., \& {Yung}, Y.~L.
	2014{\natexlab{b}}, \apj, 783, 70, \dodoi{10.1088/0004-637X/783/2/70}
	
	\bibitem[{Line \& Parmentier(2016)}]{line:2016b}
	Line, M.~R., \& Parmentier, V. 2016, Astrophys. J., 820, 78,
	\dodoi{10.3847/0004-637X/820/1/78}
	
	\bibitem[{{Line} {et~al.}(2015){Line}, {Teske}, {Burningham}, {Fortney}, \&
		{Marley}}]{line:2015}
	{Line}, M.~R., {Teske}, J., {Burningham}, B., {Fortney}, J.~J., \& {Marley},
	M.~S. 2015, \apj, 807, 183, \dodoi{10.1088/0004-637X/807/2/183}
	
	\bibitem[{{Line} {et~al.}(2012){Line}, {Zhang}, {Vasisht}, {Natraj}, {Chen}, \&
		{Yung}}]{line:2012}
	{Line}, M.~R., {Zhang}, X., {Vasisht}, G., {et~al.} 2012, \apj, 749, 93,
	\dodoi{10.1088/0004-637X/749/1/93}
	
	\bibitem[{{Line} {et~al.}(2013){Line}, {Wolf}, {Zhang}, {Knutson}, {Kammer},
		{Ellison}, {Deroo}, {Crisp}, \& {Yung}}]{line:2013}
	{Line}, M.~R., {Wolf}, A.~S., {Zhang}, X., {et~al.} 2013, \apj, 775, 137,
	\dodoi{10.1088/0004-637X/775/2/137}
	
	\bibitem[{{Line} {et~al.}(2016){Line}, {Stevenson}, {Bean}, {Desert},
		{Fortney}, {Kreidberg}, {Madhusudhan}, {Showman}, \&
		{Diamond-Lowe}}]{line:2016}
	{Line}, M.~R., {Stevenson}, K.~B., {Bean}, J., {et~al.} 2016, \aj, 152, 203,
	\dodoi{10.3847/0004-6256/152/6/203}
	
	\bibitem[{{Line} {et~al.}(2017){Line}, {Marley}, {Liu}, {Burningham}, {Morley},
		{Hinkel}, {Teske}, {Fortney}, {Freedman}, \& {Lupu}}]{line:2017}
	{Line}, M.~R., {Marley}, M.~S., {Liu}, M.~C., {et~al.} 2017, \apj, 848, 83,
	\dodoi{10.3847/1538-4357/aa7ff0}
	
	\bibitem[{{Lothringer} \& {Barman}(2019)}]{lothringer:2019}
	{Lothringer}, J.~D., \& {Barman}, T. 2019, \apj, 876, 69,
	\dodoi{10.3847/1538-4357/ab1485}
	
	\bibitem[{{Lothringer} {et~al.}(2018){Lothringer}, {Barman}, \&
		{Koskinen}}]{lothringer:2018b}
	{Lothringer}, J.~D., {Barman}, T., \& {Koskinen}, T. 2018, \apj, 866, 27,
	\dodoi{10.3847/1538-4357/aadd9e}
	
	\bibitem[{{MacDonald} \& {Madhusudhan}(2017)}]{macdonald:2017}
	{MacDonald}, R.~J., \& {Madhusudhan}, N. 2017, \mnras, 469, 1979,
	\dodoi{10.1093/mnras/stx804}
	
	\bibitem[{{Madhusudhan}(2018)}]{madhu:2018review}
	{Madhusudhan}, N. 2018, {Atmospheric Retrieval of Exoplanets} (Handbook of
	Exoplanets), 104, \dodoi{10.1007/978-3-319-55333-7_104}
	
	\bibitem[{{Madhusudhan} \& {Seager}(2009)}]{madhusudhan:2009}
	{Madhusudhan}, N., \& {Seager}, S. 2009, \apj, 707, 24,
	\dodoi{10.1088/0004-637X/707/1/24}
	
	\bibitem[{{Mansfield} {et~al.}(2018){Mansfield}, {Bean}, {Line}, {Parmentier},
		{Kreidberg}, {D{\'e}sert}, {Fortney}, {Stevenson}, {Arcangeli}, \&
		{Dragomir}}]{mansfield:2018}
	{Mansfield}, M., {Bean}, J.~L., {Line}, M.~R., {et~al.} 2018, \aj, 156, 10,
	\dodoi{10.3847/1538-3881/aac497}
	
	\bibitem[{{McKemmish} {et~al.}(2019){McKemmish}, {Masseron}, {Hoeijmakers},
		{Perez-Mesa}, {Grimm}, {Yurchenko}, \& {Tennyson}}]{mckemmish:2019}
	{McKemmish}, L.~K., {Masseron}, T., {Hoeijmakers}, H.~J., {et~al.} 2019, arXiv
	e-prints, arXiv:1905.04587.
	\newblock \doarXiv{1905.04587}
	
	\bibitem[{{Molli{\`e}re} {et~al.}(2019){Molli{\`e}re}, {Wardenier}, {van
			Boekel}, {Henning}, {Molaverdikhani}, \& {Snellen}}]{molliere:2019}
	{Molli{\`e}re}, P., {Wardenier}, J.~P., {van Boekel}, R., {et~al.} 2019, \aap,
	627, A67, \dodoi{10.1051/0004-6361/201935470}
	
	\bibitem[{{Pannekoek}(1931)}]{pannekoek:1931}
	{Pannekoek}, A. 1931, \mnras, 91, 519, \dodoi{10.1093/mnras/91.5.519}
	
	\bibitem[{{Parmentier} \& {Guillot}(2014)}]{parmentier:2014}
	{Parmentier}, V., \& {Guillot}, T. 2014, \aap, 562, A133,
	\dodoi{10.1051/0004-6361/201322342}
	
	\bibitem[{{Parmentier} {et~al.}(2013){Parmentier}, {Showman}, \&
		{Lian}}]{parmentier:2013}
	{Parmentier}, V., {Showman}, A.~P., \& {Lian}, Y. 2013, \aap, 558, A91,
	\dodoi{10.1051/0004-6361/201321132}
	
	\bibitem[{{Parmentier} {et~al.}(2018){Parmentier}, {Line}, {Bean}, {Mansfield},
		{Kreidberg}, {Lupu}, {Visscher}, {D{\'e}sert}, {Fortney}, {Deleuil},
		{Arcangeli}, {Showman}, \& {Marley}}]{parmentier:2018}
	{Parmentier}, V., {Line}, M.~R., {Bean}, J.~L., {et~al.} 2018, \aap, 617, A110,
	\dodoi{10.1051/0004-6361/201833059}
	
	\bibitem[{{Pascale} {et~al.}(2018){Pascale}, {Bezawada}, {Barstow}, {Beaulieu},
		{Bowles}, {Coud{\'e} du Foresto}, {Coustenis}, {Decin}, {Drossart},
		{Eccleston}, {Encrenaz}, {Forget}, {Griffin}, {G{\"u}del}, {Hartogh},
		{Heske}, {Lagage}, {Leconte}, {Malaguti}, {Micela}, {Middleton}, {Min},
		{Moneti}, {Morales}, {Mugnai}, {Ollivier}, {Pace}, {Papageorgiou},
		{Pilbratt}, {Puig}, {Rataj}, {Ray}, {Ribas}, {Rocchetto}, {Sarkar}, {Selsis},
		{Taylor}, {Tennyson}, {Tinetti}, {Turrini}, {Vandenbussche}, {Venot},
		{Waldmann}, {Wolkenberg}, {Wright}, {Zapatero Osorio}, \&
		{Zingales}}]{pascale:2020}
	{Pascale}, E., {Bezawada}, N., {Barstow}, J., {et~al.} 2018, in Society of
	Photo-Optical Instrumentation Engineers (SPIE) Conference Series, Vol. 10698,
	\procspie, 106980H, \dodoi{10.1117/12.2311838}
	
	\bibitem[{{Patrascu} {et~al.}(2015){Patrascu}, {Yurchenko}, \&
		{Tennyson}}]{patrascu:2015}
	{Patrascu}, A.~T., {Yurchenko}, S.~N., \& {Tennyson}, J. 2015, \mnras, 449,
	3613, \dodoi{10.1093/mnras/stv507}
	
	\bibitem[{{Pino} {et~al.}(2020){Pino}, {D{\'e}sert}, {Brogi}, {Malavolta},
		{Wyttenbach}, {Line}, {Hoeijmakers}, {Fossati}, {Bonomo}, {Nascimbeni},
		{Panwar}, {Affer}, {Benatti}, {Biazzo}, {Bignamini}, {Borsa}, {Carleo},
		{Claudi}, {Cosentino}, {Covino}, {Damasso}, {Desidera}, {Giacobbe},
		{Harutyunyan}, {Lanza}, {Leto}, {Maggio}, {Maldonado}, {Mancini}, {Micela},
		{Molinari}, {Pagano}, {Piotto}, {Poretti}, {Rainer}, {Scandariato},
		{Sozzetti}, {Allart}, {Borsato}, {Bruno}, {Di Fabrizio}, {Ehrenreich},
		{Fiorenzano}, {Frustagli}, {Lavie}, {Lovis}, {Magazz{\`u}}, {Nardiello},
		{Pedani}, \& {Smareglia}}]{pino:2020}
	{Pino}, L., {D{\'e}sert}, J.~M., {Brogi}, M., {et~al.} 2020, arXiv e-prints,
	arXiv:2004.11335.
	\newblock \doarXiv{2004.11335}
	
	\bibitem[{{Rogers} \& {Komacek}(2014)}]{rogersandtad:2014}
	{Rogers}, T.~M., \& {Komacek}, T.~D. 2014, \apj, 794, 132,
	\dodoi{10.1088/0004-637X/794/2/132}
	
	\bibitem[{{Rogers} \& {McElwaine}(2017)}]{rogers:2017b}
	{Rogers}, T.~M., \& {McElwaine}, J.~N. 2017, \apjl, 841, L26,
	\dodoi{10.3847/2041-8213/aa72da}
	
	\bibitem[{{Rothman} {et~al.}(2009){Rothman}, {Gordon}, {Barbe}, {Benner},
		{Bernath}, {Birk}, {Boudon}, {Brown}, {Campargue}, {Champion}, {Chance},
		{Coudert}, {Dana}, {Devi}, {Fally}, {Flaud}, {Gamache}, {Goldman},
		{Jacquemart}, {Kleiner}, {Lacome}, {Lafferty}, {Mandin}, {Massie},
		{Mikhailenko}, {Miller}, {Moazzen-Ahmadi}, {Naumenko}, {Nikitin}, {Orphal},
		{Perevalov}, {Perrin}, {Predoi-Cross}, {Rinsland}, {Rotger}, {{\v S}ime{\v
				c}kov{\'a}}, {Smith}, {Sung}, {Tashkun}, {Tennyson}, {Toth}, {Vandaele}, \&
		{Vander Auwera}}]{rothman:2008}
	{Rothman}, L.~S., {Gordon}, I.~E., {Barbe}, A., {et~al.} 2009, \jqsrt, 110,
	533, \dodoi{10.1016/j.jqsrt.2009.02.013}
	
	\bibitem[{{Rothman} {et~al.}(2010){Rothman}, {Gordon}, {Barber}, {Dothe},
		{Gamache}, {Goldman}, {Perevalov}, {Tashkun}, \& {Tennyson}}]{rothman:2010}
	{Rothman}, L.~S., {Gordon}, I.~E., {Barber}, R.~J., {et~al.} 2010, \jqsrt, 111,
	2139, \dodoi{10.1016/j.jqsrt.2010.05.001}
	
	\bibitem[{Schwarz(1978)}]{schwarz:1978}
	Schwarz, G. 1978, Annals of Statistics, 6, 461
	
	\bibitem[{{Schweitzer} {et~al.}(1996){Schweitzer}, {Hauschildt}, {Allard}, \&
		{Basri}}]{schweitzer:1996}
	{Schweitzer}, A., {Hauschildt}, P.~H., {Allard}, F., \& {Basri}, G. 1996,
	\mnras, 283, 821, \dodoi{10.1093/mnras/283.3.821}
	
	\bibitem[{{Skilling}(2004)}]{skilling:2004}
	{Skilling}, J. 2004, in American Institute of Physics Conference Series, Vol.
	735, American Institute of Physics Conference Series, ed. R.~{Fischer},
	R.~{Preuss}, \& U.~V. {Toussaint}, 395--405, \dodoi{10.1063/1.1835238}
	
	\bibitem[{{Ter Braak}(2006)}]{terbraak:2006}
	{Ter Braak}, C.~J.~F. 2006, Statistics and Computing, Volume 16, Issue 3, pp
	239-249, 2006, 16, \dodoi{10.1007/s11222-006-8769-1}
	
	\bibitem[{ter Braak \& Vrugt(2008)}]{terbraak:2008}
	ter Braak, C. J.~F., \& Vrugt, J.~A. 2008, Statistics and Computing, 18, 435,
	\dodoi{10.1007/s11222-008-9104-9}
	
	\bibitem[{{Waldmann} {et~al.}(2015){Waldmann}, {Tinetti}, {Rocchetto},
		{Barton}, {Yurchenko}, \& {Tennyson}}]{waldmann:2015}
	{Waldmann}, I.~P., {Tinetti}, G., {Rocchetto}, M., {et~al.} 2015, \apj, 802,
	107, \dodoi{10.1088/0004-637X/802/2/107}
	
	\bibitem[{{Welbanks} \& {Madhusudhan}(2019)}]{welbanks:2019}
	{Welbanks}, L., \& {Madhusudhan}, N. 2019, \aj, 157, 206,
	\dodoi{10.3847/1538-3881/ab14de}
	
	\bibitem[{{Wildt}(1939)}]{wildt:1939}
	{Wildt}, R. 1939, \apj, 90, 611, \dodoi{10.1086/144125}
	
	\bibitem[{{Yurchenko} {et~al.}(2018){Yurchenko}, {Bond}, {Gorman}, {Lodi},
		{McKemmish}, {Nunn}, {Shah}, \& {Tennyson}}]{yurchenko:2018}
	{Yurchenko}, S.~N., {Bond}, W., {Gorman}, M.~N., {et~al.} 2018, \mnras, 478,
	270, \dodoi{10.1093/mnras/sty939}
	
	\bibitem[{{Zahnle} {et~al.}(2009){Zahnle}, {Marley}, {Freedman}, {Lodders}, \&
		{Fortney}}]{zahnle:2009b}
	{Zahnle}, K., {Marley}, M.~S., {Freedman}, R.~S., {Lodders}, K., \& {Fortney},
	J.~J. 2009, \apjl, 701, L20, \dodoi{10.1088/0004-637X/701/1/L20}
	
\end{thebibliography}

\end{document}